\pdfoutput=1
\def\paperTitle{PoTAS: Post-hoc Tail Alignment of Similarity Scores\\for Embedding Compatibility with Mixed Databases}

\def\authorBlock{
Jaeseok Byun \qquad Joonyoung Chang \qquad Taesup Moon\thanks{Corresponding author.} \\
Seoul National University \\
{\tt\small \{wotjr3868, jjoon0, tsmoon\}@snu.ac.kr}
}

\documentclass[10pt,twocolumn,letterpaper]{article}

\usepackage{times}
\usepackage{xspace}
\usepackage[dvipsnames]{xcolor}
\usepackage{graphicx}
\usepackage{amsmath}
\usepackage{amssymb}
\usepackage{booktabs}
\usepackage[numbers,sort&compress]{natbib}
\usepackage{silence}
\usepackage{etoolbox}
\usepackage[format=plain,labelformat=simple,labelsep=period,font=small,compatibility=false]{caption}
\usepackage[font=footnotesize,skip=3pt,subrefformat=parens]{subcaption}
\usepackage[hyphens]{url}
\makeatletter
\font\elvbf = ptmb scaled 1100
\def\@maketitle{
   \newpage
   \null
   \vskip .2in
   \begin{center}
      {\Large \bf \@title \par}
      \vspace*{16pt}{
        \large
        \lineskip .5em
        \begin{tabular}[t]{c}
            \@author
        \end{tabular}
        \par
      }
      \vskip .5em
      \vspace*{8pt}
   \end{center}
}
\def\abstract{%
   \centerline{\large\bf Abstract}%
   \vspace*{12pt}\noindent%
   \it\ignorespaces%
}

\def\pp@section{\@startsection {section}{1}{\z@}
   {-10pt plus -2pt minus -2pt}{7pt} {\large\bf}}
\def\pp@ssect#1{\pp@section*{#1}}
\def\pp@sect#1{\pp@section{\texorpdfstring{\hskip -1em.~}{}#1}}
\def\section{\@ifstar\pp@ssect\pp@sect}
\def\pp@subsection{\@startsection {subsection}{2}{\z@}
   {-8pt plus -2pt minus -2pt}{5pt} {\elvbf}}
\def\pp@ssubsect#1{\pp@subsection*{#1}}
\def\pp@subsect#1{\pp@subsection{\texorpdfstring{\hskip -1em.~}{}#1}}
\def\subsection{\@ifstar\pp@ssubsect\pp@subsect}

\DeclareRobustCommand\onedot{\futurelet\@let@token\@onedot}
\def\@onedot{\ifx\@let@token.\else.\null\fi\xspace}
\def\eg{\emph{e.g}\onedot}
\def\ie{\emph{i.e}\onedot}
\makeatother
\AtEndPreamble{
    \usepackage[capitalize]{cleveref}
    \crefname{section}{Sec.}{Secs.}
    \Crefname{section}{Section}{Sections}
    \Crefname{table}{Table}{Tables}
    \crefname{table}{Tab.}{Tabs.}
}
\usepackage[shortlabels,inline]{enumitem}
\setlist[itemize]{noitemsep,leftmargin=*,topsep=0em}
\setlist[enumerate]{noitemsep,leftmargin=*,topsep=0em}

\usepackage{graphicx}	
\usepackage{amsmath}	
\usepackage{amssymb}
\usepackage{pifont}	
\usepackage{booktabs}
\usepackage{times}
\usepackage{microtype}
\usepackage{epsfig}
\usepackage{caption}
\usepackage{float}
\usepackage{color, colortbl}
\usepackage{stfloats}
\usepackage{enumitem}
\usepackage{tabularx}
\usepackage{xstring}
\usepackage{multirow}
\usepackage{xspace}
\usepackage{url}
\usepackage{subcaption}
\usepackage{xcolor}
\usepackage[hang,flushmargin]{footmisc}

\newcommand{\nbf}[1]{{\noindent \textbf{#1.}}}

\definecolor{gaincolor}{RGB}{0,120,80}
\definecolor{losscolor}{RGB}{180,40,40}
\definecolor{oursgray}{gray}{0.92}
\newcommand{\cmark}{\ding{51}}
\newcommand{\xmark}{\ding{55}}

\newcommand{\sdUpgMax}{0.003\xspace}

\newcommand{\tmc}{TMC\xspace}
\newcommand{\thn}{THN\xspace}
\newcommand{\ours}{PoTAS\xspace}
\newcommand{\rmnaive}{\ensuremath{\mathrm{RM}_{\text{old}}}\xspace}
\newcommand{\rmnaiveh}{\ensuremath{\mathrm{RM}_{\text{bc}}}\xspace}

\definecolor{linkblue}{rgb}{0.21,0.49,0.74}
\usepackage[pagebackref,breaklinks,colorlinks,allcolors=linkblue]{hyperref}
\hypersetup{pdftitle={PoTAS: Post-hoc Tail Alignment of Similarity Scores for Embedding Compatibility with Mixed Databases},pdfauthor={Jaeseok Byun, Joonyoung Chang, Taesup Moon}}
\usepackage[capitalize]{cleveref}
\crefname{section}{Sec.}{Secs.}
\crefname{table}{Table}{Tables}
\crefname{figure}{Fig.}{Figs.}

\crefname{appendix}{App.}{Apps.}

\begin{document}
\title{\paperTitle}
\author{\authorBlock}
\maketitle

\begin{abstract}
Upgrading embedding models in deployed retrieval systems often requires costly re-embedding of the entire gallery due to embedding incompatibility.
Backward-compatible methods enable new queries to retrieve old embeddings, but gradual migration results in mixed databases requiring retrieval results from old and new embeddings to be combined.
We find that mismatched similarity score distributions make this combination challenging in realistic cross-model, cross-modal upgrades.
To address this, we propose \ours, a post-hoc framework that aligns the high-score tails governing retrieval result merging without retraining either embedding backbone.
It combines \emph{tail-matched calibration} (\tmc), which aligns score tails at partition-size-aware depths without training or reference data, with \emph{tail-focused hard-negative refinement} (\thn), which refines the backward adapter using high-scoring negatives.
Across image--text and visual document retrieval, diverse model upgrades, and multiple adapter-based backward-compatible methods, \ours is the only approach that, on average, consistently outperforms the old model from the onset of migration, surpassing all baselines by substantial margins with minimal computational overhead.

\end{abstract}
\section{Introduction}
\label{sec:intro}

\begin{figure}[t]
  \centering
  \includegraphics[width=\linewidth]{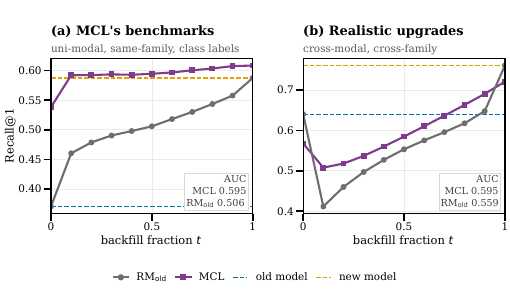}
  \caption{\textbf{Rank merging and MCL on MCL's benchmarks versus realistic upgrades.} Recall@1 over the backfill
  (dashed: old and new model alone; box: AUC). (a)~Image-to-image retrieval with class labels,
  mean of seven setups of MCL~\cite{seo2025mcl} on ImageNet-1K~\cite{deng2009imagenet}, CIFAR-100~\cite{krizhevsky2009cifar} and Places-365~\cite{zhou2017places}, reproduced as described in the appendix: on average, both rise steadily
  as migration proceeds. (b)~Cross-modal retrieval with few answers per query, four-upgrade mean on COCO~\cite{lin2014coco}, Flickr30K~\cite{plummer2015flickr} and Urban1k~\cite{zhang2024longclip}: both drop sharply at the start and recover slowly.}
  \label{fig:backfill}
\end{figure}

Embedding models have become fundamental components of modern information retrieval systems, enabling
semantic search over text~\cite{li2023gte} as well as across modalities such as images and
video~\cite{radford2021clip,jia2021align,luo2021clip4clip}.
Their role has grown further with the rise of large language models (LLMs).
For example, retrieval-augmented generation (RAG) grounds LLM outputs in documents retrieved by embedding models~\cite{lewis2020rag}, while LLM agents use embedding-based memory retrieval to store and recall past experiences~\cite{park2023generative,packer2023memgpt}.
More recently, universal multimodal embedders built on multimodal LLMs (MLLMs) have emerged to support diverse modalities and retrieval tasks within a unified embedding space~\cite{jiang2024e5v,jiang2025vlm2vec,zhang2025gme,gu2025unime}.

Deployed retrieval systems often upgrade their embedding models to meet evolving demands.
However, new models typically produce embeddings incompatible with the old database, requiring the entire gallery to be re-embedded with the new model, a costly process known as \emph{backfilling}~\cite{shen2020bct}, particularly at billion-scale or for multimodal data.
Backward-compatible training (BCT)~\cite{shen2020bct,meng2021lce,hu2022lbce} addresses this by enabling new query embeddings to be compared directly with old gallery embeddings, while XBT~\cite{jang2025xbt} extends this approach to vision--language models. 
However, these methods require training from scratch or modifying the new model's backbone, making them unsuitable for off-the-shelf or API-served foundation models. Adapter-based approaches instead freeze both models and learn a lightweight \emph{backward-compatible adapter} (backward adapter for short) that maps the new embedding space to the old one~\cite{vejendla2025drift,byun2026mkp}.
MKP-Adapter~\cite{byun2026mkp} preserves the new embedding space's structure, enabling cross-model, cross-modal upgrades such as from CLIP~\cite{radford2021clip} to the Qwen2-VL-based~\cite{wang2024qwen2vl} GME~\cite{zhang2025gme}.

However, these approaches assume a static gallery, whereas real-world upgrades often involve gradual migration, resulting in a \emph{mixed database} of old and new embeddings.
Forward-compatible approaches~\cite{ramanujan2022fct,jaeckle2023fastfill} transform old embeddings into the new space, but suffer from a computational bottleneck as the entire old gallery must be transformed before retrieval can begin.
In contrast, metric-compatible contrastive learning (MCL)~\cite{seo2025mcl} directly addresses retrieval over a mixed database in a unimodal setting.
MCL first presents a plain \emph{rank merge} (\rmnaive, gray in \cref{fig:backfill}), which directly combines the two ranked lists by their raw similarity scores despite differences between the retrieval systems, and then adds joint merge training on top of it (MCL, purple in \cref{fig:backfill}) to improve score comparability across the two systems.

The MCL study reports that even plain rank merging improves retrieval performance from the onset of backfilling.
We attribute this largely to two favorable benchmark properties.
First, the benchmarks involve relatively mild, unimodal model upgrades (\eg ResNet-18$\rightarrow$ResNet-50), so the two models produce similar similarity score distributions (\cref{fig:distributions}a).
Second, class-level relevance provides multiple positives per query, typically distributed across both migrated and unmigrated portions of the gallery, making rank merging less sensitive to which retrieval system is favored.
These advantages need not hold in more realistic cross-model, cross-modal retrieval settings, where similarity score distributions differ substantially and instance-level relevance typically yields few positives in only one gallery partition.
Consequently, both plain rank merging and MCL drop sharply below the old model as soon as migration begins and recover only slowly (\cref{fig:backfill}b).

We believe these results suggest that MCL’s rank merge and joint merge training are insufficient for effective retrieval when the two systems’ similarity score distributions differ substantially, highlighting the need for explicit score alignment across systems.
A natural approach is to fit a fixed mapping between the two systems’ scores on a reference set embedded by both models. However, this mapping may not generalize to the deployed data, particularly as queries and partitions change. Calibrating directly on the deployed data is also challenging because the mixed gallery provides no paired scores: each gallery item is scored by only one of the two systems.
To address this, we propose \ours, a post-hoc framework that aligns the high-score tails that decide the merge, with two complementary components:
(1) \emph{Tail-matched calibration} (\tmc) aligns tail score distributions at inference time using the current gallery partitions and depths proportional to their sizes, without training or reference data. 
(2) \emph{Tail-focused hard-negative refinement} (\thn) further refines the trained backward adapter by emphasizing high-scoring negatives.
We evaluate on image--text (COCO~\cite{lin2014coco}, Flickr30K~\cite{plummer2015flickr}, Urban1k~\cite{zhang2024longclip}) and visual document retrieval (ViDoRe v1/v2~\cite{faysse2025colpali,mace2025vidorev2}), model upgrades (CLIP~\cite{radford2021clip}$\rightarrow$MLLM~\cite{gu2025unime,zhang2025gme}, CLIP$\rightarrow$CLIP, MLLM~\cite{jiang2024e5v}$\rightarrow$MLLM) and three adapter-based BCT methods (MKP-Adapter~\cite{byun2026mkp}, Drift-Adapter~\cite{vejendla2025drift}, RQT~\cite{seo2025mcl}): \ours consistently improves every method and upgrade.
Our contributions are:
\begin{itemize}[leftmargin=*,itemsep=1pt,topsep=2pt]
  \item  We show that existing methods for mixed-database retrieval are unreliable in realistic cross-model, cross-modal upgrades, identifying similarity score mismatch as a key cause (\cref{sec:motivation}).
\item We propose \ours, a computationally efficient post-hoc framework that aligns the high-score tails governing merge decisions through training-free inference-time calibration and lightweight refinement of the backward adapter (\cref{sec:calibration,sec:merge_training}).
\item Across retrieval tasks, model upgrades, and adapter-based BCT methods, \ours is the only method that, on average, consistently outperforms the old model from the onset of migration, surpassing all baselines by large margins. In contrast, existing baselines often perform worse than the old model for much of the migration, limiting their practical applicability (\cref{sec:exp}).
\end{itemize}
\section{Related Work}
\label{sec:related}

\nbf{Embedding compatibility}
Forward-compatible methods~\cite{ramanujan2022fct,jaeckle2023fastfill} transform the whole old gallery into the new
space before the new model can serve, while backward-compatible
training~\cite{shen2020bct,meng2021lce,hu2022lbce,zhang2022hotrefresh,su2022bict,jang2025xbt} and backward
adapters~\cite{seo2025mcl,vejendla2025drift,byun2026mkp} make new queries comparable with an entirely old gallery.
During the backfill, the gallery is a mixed database of old and new embeddings. MCL~\cite{seo2025mcl}
rank-merges the two systems' lists and adds joint merge training so that their scores become comparable, on
uni-modal, class-labeled benchmarks. \ours instead starts post hoc from a backward adapter and aligns the two
systems' scores at inference time, on the tails that decide the merge.

\nbf{Merging ranked lists}
Reciprocal rank fusion (RRF)~\cite{cormack2009rrf} combines ranked lists from multiple search engines based on ranks alone, whereas Metasearch~\cite{montague2001normalization} normalizes their scores (\eg min-max or z-score normalization) before summation.
In speaker verification, AS-norm~\cite{matejka2017asnorm} normalizes scores using the mean and standard deviation computed from a cohort of the most similar impostors.
These methods rely on information derived independently from each ranked list.
However, mixed-database retrieval requires ranking disjoint partitions of varying sizes, where larger partitions naturally tend to yield higher top-1 scores.
Per-list normalization obscures this size-dependent difference, with min-max normalization and RRF even assigning identical values to the top-ranked items from both partitions.
Our \tmc maps the new scores onto the old scale with tail statistics taken at depths proportional to the partition
sizes, so that both are read at the same gallery quantile.

\section{Problem Setup and Preliminaries}

\nbf{Problem setup}
Let $f_{\text{old}}$ and $f_{\text{new}}$ be the frozen old and new embedding models, and
$\mathbf{x}_i^{\text{old}}\in\mathbb{R}^{d_{\text{old}}}$ and $\mathbf{x}_i^{\text{new}}\in\mathbb{R}^{d_{\text{new}}}$ the
$\ell_2$-normalized embeddings of a sample $s_i$. During backfilling the gallery $\mathcal{G}$ is re-embedded gradually: at
backfill fraction $t\in[0,1]$, the migrated part $\mathcal{G}^{\text{new}}_t$ ($\lfloor t|\mathcal{G}|\rfloor$ items) holds
new embeddings and the rest, $\mathcal{G}^{\text{old}}_t$, old ones. For the $i$-th query, the \emph{new system} (the new model) scores the
migrated items and the \emph{backward system} (old-space queries on old embeddings) the rest:
\begin{equation}
  S_{ij}=
  \begin{cases}
    S^{\text{new}}_{ij}=\cos\big(\mathbf{x}_i^{\text{new}},\mathbf{x}_j^{\text{new}}\big) & s_j\in\mathcal{G}^{\text{new}}_t,\\[2pt]
    S^{\text{old}}_{ij}=\cos\big(\mathbf{q}_i,\mathbf{x}_j^{\text{old}}\big) & s_j\in\mathcal{G}^{\text{old}}_t,
  \end{cases}
\end{equation}
where $\mathbf{q}_i$ is the $i$-th query in the old space.

\nbf{Rank merge from MCL~\cite{seo2025mcl}}
\label{sec:mcl}
MCL first introduced the \emph{rank merge} (RM) for retrieval from a mixed database: each system retrieves from its
own partition, and the two lists are interleaved by their raw scores, without re-scaling. The merged ranking thus
sorts all items of $\mathcal{G}$ by $S_{ij}$, and its top-1 item $\hat{j}_i=\arg\max_{j\in\mathcal{G}}S_{ij}$ comes
from the partition whose own top-1 score is higher:
\begin{equation}
  \hat{j}_i\in\mathcal{G}^{\text{new}}_t\iff\max_{j\in\mathcal{G}^{\text{new}}_t}S^{\text{new}}_{ij}>\max_{j\in\mathcal{G}^{\text{old}}_t}S^{\text{old}}_{ij}.
  \label{eq:rm}
\end{equation}
The rank merge takes two forms, which differ only in $\mathbf{q}_i$. \rmnaive uses the original old embeddings,
$\mathbf{q}_i=\mathbf{x}_i^{\text{old}}$. \rmnaiveh uses a backward-compatible adapter
$h:\mathbb{R}^{d_{\text{new}}}\rightarrow\mathbb{R}^{d_{\text{old}}}$~\cite{byun2026mkp},
$\mathbf{q}_i=h(\mathbf{x}_i^{\text{new}})$, so the backward system needs only $f_{\text{new}}$ at inference; as $h$ we use the adapters of \cref{sec:baselines}.

\nbf{Joint merge training from MCL}
MCL further trains $h$ together with a map $\rho$ in the new space, applied to everything the new system
serves, with a supervised contrastive loss~\cite{khosla2020supcon} whose
denominator also contains the other system's negatives. Within a training batch, both systems score every item $j$ for
anchor $i$ as they would at inference time: the backward system with the adapted query against $j$'s old embedding,
$S^{\text{old}}_{ij}=\cos\big(h(\mathbf{x}_i^{\text{new}}),\mathbf{x}_j^{\text{old}}\big)$, and the new system with
$S^{\text{new}}_{ij}=\cos\big(\rho(\mathbf{x}_i^{\text{new}}),\rho(\mathbf{x}_j^{\text{new}})\big)$. With class labels
defining the positives $\mathcal{P}_i$ and negatives $\mathcal{N}_i$ of anchor $i$,
\begin{equation}
\begin{aligned}
  \mathcal{L}_{\text{MCL}}=\;&-\log\frac{\sum_{j\in\mathcal{P}_i}e^{S^{\text{old}}_{ij}/\tau}}
  {\sum_{j\in\mathcal{P}_i}e^{S^{\text{old}}_{ij}/\tau}+\sum_{j\in\mathcal{N}_i}\big(e^{S^{\text{old}}_{ij}/\tau}+e^{S^{\text{new}}_{ij}/\tau}\big)}\\
  &-\log\frac{\sum_{j\in\mathcal{P}_i}e^{S^{\text{new}}_{ij}/\tau}}
  {\sum_{j\in\mathcal{P}_i}e^{S^{\text{new}}_{ij}/\tau}+\sum_{j\in\mathcal{N}_i}\big(e^{S^{\text{new}}_{ij}/\tau}+e^{S^{\text{old}}_{ij}/\tau}\big)}.
\end{aligned}
\label{eq:mcl}
\end{equation}
\section{Motivational analysis}
\label{sec:motivation}

\begin{figure*}[t]
  \centering
  \includegraphics[width=\linewidth]{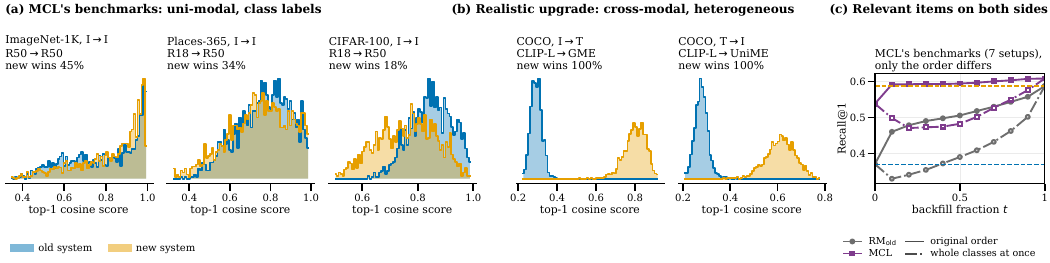}
  \caption{\textbf{What separates MCL's benchmarks from realistic upgrades.}
(a,b) Distributions of per-query top-1 scores from the old and new systems; ``new wins'' denotes the fraction of queries with a higher new-system score.
(a) Image-to-image retrieval on three MCL benchmarks (ImageNet-1K R50$\rightarrow$R50, Places-365 and CIFAR-100 R18$\rightarrow$R50; \cref{fig:backfill}a): distributions overlap, with neither system consistently scoring higher.
(b) COCO image-to-text (CLIP-L$\rightarrow$GME) and text-to-image (CLIP-L$\rightarrow$UniME): distributions are disjoint, with the new system scoring higher for every query.
(c) Mean Recall@1 over backfilling across seven MCL setups, comparing MCL's original migration order (solid), which distributes positives across both partitions, with random class-wise migration (dash-dot), which keeps them in one partition.
Models, labels, queries, and galleries are identical; only migration order differs.
Dashed: old and new models alone. Per-setup results are in the appendix}

  \label{fig:distributions}
\end{figure*}

\nbf{MCL's setting is mild}
The model upgrades considered by MCL~\cite{seo2025mcl} are relatively mild, involving either retraining the same architecture on more classes or upgrading within a model family (\eg ResNet-18$\rightarrow$ResNet-50), with one exception involving ViT-B/32.
All experiments use unimodal image retrieval with class-level relevance, yielding 49 (ImageNet-1K) to 99 (CIFAR-100 and Places-365) relevant gallery items per query.
Across seven reproduced settings (ImageNet-1K~\cite{deng2009imagenet}, CIFAR-100, and Places-365, using ResNet~\cite{he2016resnet} and ViT models; details in the appendix), \rmnaive outperforms the old model on average from the onset of migration, while MCL further improves performance (\cref{fig:backfill}a).

\nbf{Realistic scenarios break both}
Realistic retrieval scenarios often involve cross-modal retrieval, heterogeneous model upgrades, and instance-level relevance with few positives per query.
In these settings, \rmnaive drops sharply as migration begins, while MCL, reproduced under the same conditions (details in the appendix), remains below the old model until late in migration (\cref{fig:backfill}b).
Two key properties explain this gap.

\nbf{(1) Mismatched similarity score distributions}
\Cref{fig:distributions} compares the distributions of per-query top-1 scores from both systems, which directly determine the merge.
On MCL's benchmarks, the two distributions largely overlap in most cases (\cref{fig:distributions}a).
In contrast, for cross-modal retrieval with CLIP-L\(\rightarrow\)GME and CLIP-L\(\rightarrow\)UniME, the distributions are clearly separated, with the new system consistently producing higher scores (\cref{fig:distributions}b).
In this case, plain rank merging selects only migrated items regardless of their relevance, explaining the sharp performance drop as migration begins.
MCL's joint merge training also struggles when applied directly to such mismatched distributions.
We hypothesize that closing the large score gap conflicts with preserving retrieval quality, pulling the backward adapter away from the old embedding space and degrading backward retrieval below the old model (\cref{sec:exp}).
Modality and model-family differences can both contribute to this mismatch, with their effects compounding (see the appendix).
Conversely, plain rank merging and MCL work well when the two score distributions are similar, pointing to score mismatch as a key cause.

\nbf{(2) Multiple positives on both sides in MCL's setting}
In MCL's benchmarks, class-level relevance gives each query multiple positives (49--99), typically distributed across both portions of the mixed gallery.
Thus, favoring the wrong system often incurs little penalty, even when their score distributions differ (\eg CIFAR-100 R18$\rightarrow$ViT-B/32; see the appendix).
In contrast, cross-modal retrieval has far fewer positives (one for text-to-image and five for image-to-text), making incorrect system selection more costly.
To isolate this effect, we retain MCL's models, labels, queries, and gallery but migrate entire classes together, placing all positives of each query in a single partition.
Under this setting, both MCL and \rmnaive exhibit a sharp performance drop at the onset of backfilling (\cref{fig:distributions}c), resembling the behavior observed in realistic upgrades (\cref{fig:backfill}b; see the appendix).
This suggests that MCL's gains on its original benchmarks largely benefit from positives being present in both partitions. 
\section{Post-hoc tail alignment (\ours)}
\label{sec:method}
Motivated by the score-scale mismatch identified in \cref{sec:motivation}, we propose \ours, a post-hoc framework for reliable rank merging over mixed databases.
Our key idea is to focus on the upper tails of the similarity score distributions, where rank merging decisions are made.
Our framework consists of two complementary components: \tmc (\cref{sec:calibration}) calibrates the upper-tail scores at inference time, while \thn (\cref{sec:merge_training}) further refines the backward adapter by focusing training on the hardest negatives.
Note that \ours keeps both embedding backbones $f_{\text{old}}$ and $f_{\text{new}}$ frozen.

\subsection{Naive approach}
\label{sec:calib_motivation}
At backfill fraction $t$, rank merging in \cref{eq:rm} compares the highest similarity scores from the migrated gallery $\mathcal{G}^{\text{new}}_t$ ($tG$ items) and the remaining gallery $\mathcal{G}^{\text{old}}_t$ ($(1-t)G$ items).
To make these scores comparable, we consider an affine transformation of the new-system scores:
\begin{equation}
  S^{\text{new}}_{ij}\;\mapsto\;a_t\,S^{\text{new}}_{ij}+b_{i,t},
  \label{eq:affine}
\end{equation}
so that the transformed score becomes comparable with the backward-system score
$S^{\text{old}}_{ij}=\cos\big(h(\mathbf{x}_i^{\text{new}}),\mathbf{x}_j^{\text{old}}\big)$, computed with the backward adapter $h$.

A naive approach is a \emph{global map}, which uses fixed parameters $a_t=a$ and $b_{i,t}=b$, fitted once by linear
regression on the paired scores of a reference set embedded by both models. However, a map fitted offline may not generalize to deployment queries or adapt to the changing
gallery partitions, whose sizes affect the top scores.

A natural alternative is to calibrate at inference time, on the test data itself.
However, a mixed database provides no item-wise paired scores: each gallery item belongs to either the new or old partition and is scored by only one system.
Consequently, unlike reference-set calibration, we cannot fit a map through paired-score regression.
Instead, we estimate the mapping by matching the two systems' observed score distributions, without item-wise correspondence.

\subsection{Tail-matched calibration (\tmc)}
\label{sec:calibration}

To this end, we devise \tmc, which calibrates the two systems at inference time by matching the upper tails of their unpaired score distributions, taken at depths proportional to the partition sizes.
It relies only on the current query's scores and running statistics from previously served queries.
\tmc estimates a shared scale $a_t$ and intercept $\bar b_t$ from pooled upper tails (i), adjusts tail depths to account for partition sizes (ii), and derives a query-specific intercept $b_{i,t}$ (iii).

\nbf{(i) Tail-focused score alignment}
Since rank merging compares the highest scores from the two partitions, calibration should focus on their upper tails rather than the full score distributions.
We therefore estimate calibration parameters from the upper tails.
For query $i$, let the vectors $\mathbf{u}^{\text{new}}_{i,k}=\operatorname{top}_k\{S^{\text{new}}_{ij}\}_{j\in\mathcal{G}^{\text{new}}_t}$ and $\mathbf{u}^{\text{old}}_{i,k}=\operatorname{top}_k\{S^{\text{old}}_{ij}\}_{j\in\mathcal{G}^{\text{old}}_t}$ in $\mathbb{R}^k$ hold the top-$k$ scores of each system over its own partition; their entries belong to different items and are never paired.
Similar to Metasearch~\cite{montague2001normalization}, we match the two systems' scores by mean and standard deviation; unlike Metasearch, however, we use only these upper tails, not whole score lists, pooled over previously served queries $\mathcal{Q}$:
\begin{equation}
\begin{aligned}
  a_t&=\operatorname{std}\big(\mathcal{U}^{\text{old}}_t\big)\,/\,\operatorname{std}\big(\mathcal{U}^{\text{new}}_t\big),\\
  \bar b_t&=\operatorname{mean}\big(\mathcal{U}^{\text{old}}_t\big)-a_t\operatorname{mean}\big(\mathcal{U}^{\text{new}}_t\big),
\end{aligned}
  \label{eq:pooled}
\end{equation}
where $\mathcal{U}^{\text{old}}_t=\{\mathbf{u}^{\text{old}}_{i,k_o}\}_{i\in\mathcal{Q}}$ and
$\mathcal{U}^{\text{new}}_t=\{\mathbf{u}^{\text{new}}_{i,k_n}\}_{i\in\mathcal{Q}}$ pool the entries of these vectors, and \cref{eq:pooled} matches the two pooled distributions by mean and standard deviation ($a_t$ and $\bar b_t$ are scalars); the depths $k_n,k_o$ follow in (ii).

\tmc maintains running statistics of each system's pooled tail scores, storing only their count, sum, and sum of squares to compute the mean and standard deviation in \cref{eq:pooled}.
These statistics are updated after each query without storing past queries or individual scores; the first query uses its own tails.

\nbf{(ii) Partition-aware quantile matching}
To align the upper tails, we must also determine how many top scores to sample from each partition.
Even when both systems have identical score distributions, a larger partition tends to yield higher maximum scores simply because it contains more items.
Calibration should not mistake this effect for a difference in score scale.
Under uniform random migration, the larger partition is also more likely to contain relevant items.
However, taking the same number of top-$k$ scores from both partitions samples different quantiles.
For example, at $t=0.1$, the top-$k$ scores cover a much larger fraction of the migrated partition, extending further down its score distribution than in the remaining partition.

To compare the same quantiles, we scale the tail depths with partition sizes:
\begin{equation}
  k_n=\operatorname{round}\!\big(k\,t\big),\qquad
  k_o=\operatorname{round}\!\big(k\,(1-t)\big),
  \label{eq:depth}
\end{equation}
with a minimum depth of 2.
Under random migration, both tails correspond to approximately the same top-$k/G$ fraction of their respective partitions, as if each system scored the full gallery.
Thus, calibration captures the systems' score-scale differences while keeping the natural partition-size effect.

\nbf{(iii) Query-adaptive intercept}
Although the scale $a_t$ and pooled intercept $\bar b_t$ are shared across queries, the score gap between systems may vary by query.
To account for this variation, we estimate a query-specific offset from depth-matched tails and combine it with the pooled intercept:
\begin{equation}
  b_{i,t}=\beta\Big(\mu\big(\mathbf{u}^{\text{old}}_{i,k'_o}\big)-a_t\,\mu\big(\mathbf{u}^{\text{new}}_{i,k'_n}\big)\Big)+(1-\beta)\,\bar b_t,
\end{equation}
where $\mu(\cdot)$ is the mean, $k'_n,k'_o$ follow \cref{eq:depth} with a larger total depth $k'>k$ to reduce noise, and $\beta\in[0,1]$ shrinks the offset toward $\bar b_t$ (see the appendix).
We keep the scale $a_t$ shared across queries, as estimating it from a single query's limited tail scores would be unreliable.

\subsection{Tail-focused hard-negative refinement (\thn)}
\label{sec:merge_training}

While \tmc substantially improves rank merging through inference-time score calibration, it leaves the backward adapter $h$ unchanged.
We therefore introduce \thn to further adapt an existing backward adapter $h_0$ for mixed-database retrieval, while keeping both embedding backbones frozen.
Following the same upper-tail intuition as \tmc, \thn focuses training on the hardest negatives that most strongly affect rank merging.
Following MCL~\cite{seo2025mcl}, we additionally train a transformation $\rho$ of the new embeddings, but anchor it to the identity to preserve the new embedding space.

\nbf{Tail-focused objective}
For each mini-batch of $B$ image--text pairs, we add to the in-batch gallery a negative bank of $N_{\text{bank}}$ randomly sampled training items.
To simulate a mixed database during migration, we sample $t\sim\mathcal{U}(0.1,0.9)$ and migrate each gallery item with probability $t$, yielding
\begin{equation}
  M_{ij}=
  \begin{cases}
    S^{\text{new}}_{ij} & m_j=1,\\
    S^{\text{old}}_{ij} & m_j=0,
  \end{cases}
\end{equation}
where $m_j\sim\mathrm{Bernoulli}(t)$, $S^{\text{old}}_{ij}$ uses the current backward adapter $h$, and
$S^{\text{new}}_{ij}=\cos\big(\rho(\mathbf{x}_i^{\text{new}}),\rho(\mathbf{x}_j^{\text{new}})\big)$.

We optimize the score gap between each relevant item and its hardest negative:
\begin{equation}
  \mathcal{L}_{\text{tail}}=-\frac1B\sum_{i=1}^{B}\log\sigma\!\left(\frac{M_{ii}-\max_{j\neq i}M_{ij}}{T}\right),
  \label{eq:merge}
\end{equation}
applied in both retrieval directions with temperature $T$.
Unlike conventional contrastive learning, which aggregates negatives through log-sum-exp, \thn concentrates gradients on the relevant item and its hardest negative.

\nbf{Anchored $\rho$}
We additionally apply a transformation $\rho$ to new embeddings (both queries and migrated items).
Unlike MCL's $\rho$, which is trained from scratch, we use a residual linear map
$\rho(\mathbf{x})=\operatorname{normalize}(\mathbf{x}+W\mathbf{x})$,
initialized with $W=0$ to preserve the new embedding space.
We further regularize $\rho$ toward the identity transformation to limit unnecessary changes to the original embeddings.

Our final post-hoc training objective jointly optimizes the backward adapter $h$, initialized from $h_0$, and $\rho$:
\begin{equation}
  \mathcal{L}_{\text{THN}}=\mathcal{L}_{\text{tail}}+\lambda\big(1-\cos(\rho(\mathbf{x}^{\text{new}}),\mathbf{x}^{\text{new}})\big),
  \label{eq:total}
\end{equation}
where $\lambda$ weights the batch-averaged regularization term.

\nbf{Inference}
The trained $h$ and $\rho$ from \thn are used at inference, where \tmc calibrates $S^{\text{new}}$ for each query via \cref{eq:affine}, followed by rank merging across the two partitions.
\section{Experiments}
\label{sec:exp}

\subsection{Setup}
\begin{figure*}[t]
  \centering
  \includegraphics[width=\linewidth]{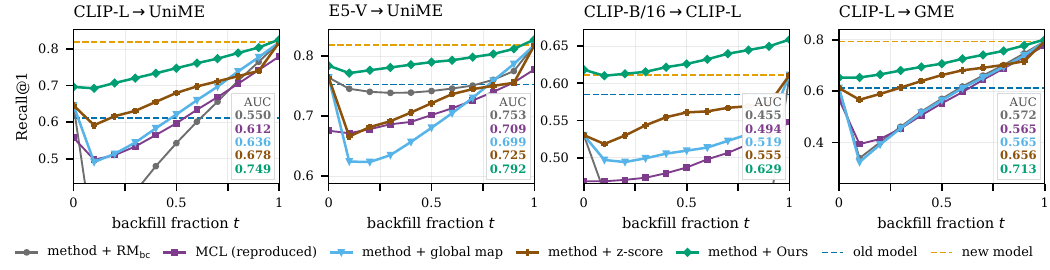}
  \caption{\textbf{From \rmnaiveh to tail-focused calibration and training.}
  Recall@1 over the backfill on the four upgrades, averaged over the three backward-compatible adapters of \cref{tab:main} and
  over COCO, Flickr30K and Urban1k in both directions. Method + \rmnaiveh: each method with plain rank merging;
  + global map and + z-score: the same adapter with the score normalizations of \cref{sec:baselines} (z-score:
  Metasearch~\cite{montague2001normalization}); method + Ours: its adapter trained with \thn and merged with \tmc.
  MCL~\cite{seo2025mcl}: merged by plain rank merging. $\downarrow$: entirely below the axis. Dashed: old and new model alone. Same runs as \cref{tab:main}; box: AUC, the mean of its columns over the adapters. Each adapter is shown separately in \cref{fig:main_curves}.}
  \label{fig:calibration_curves}
\end{figure*}

\nbf{Model upgrades}
We study four model upgrades spanning different levels of heterogeneity (\cref{sec:motivation}):
CLIP-B/16$\rightarrow$CLIP-L/14~\cite{radford2021clip}, a size upgrade within the same model family;
E5-V~\cite{jiang2024e5v}$\rightarrow$UniME-7B~\cite{gu2025unime}, between MLLM embedders built on LLaVA-1.6~\cite{liu2024llavanext};
and CLIP-L/14$\rightarrow$UniME-7B and CLIP-L/14$\rightarrow$GME-7B~\cite{zhang2025gme} (on Qwen2-VL~\cite{wang2024qwen2vl}), from a contrastive dual encoder to MLLM embedders.
For visual document retrieval, we use CLIP-L/14$\rightarrow$GME-7B, following~\cite{byun2026mkp}: among our new models, only GME suits visual documents. Both encoders stay frozen.

\nbf{Data}
Following~\cite{byun2026mkp}, adapters are trained on pre-extracted embeddings of LLaVA-LCS-558K
image--caption pairs~\cite{liu2023llava} (images, captions, and image--caption composites); the backward adapters use a fixed
random subset of 500K embeddings, and \thn training samples 200K image--caption pairs.
We evaluate on the COCO 5K~\cite{lin2014coco}, Flickr30K 1K~\cite{plummer2015flickr} and
Urban1k~\cite{zhang2024longclip} test sets in both directions (image$\rightarrow$text and text$\rightarrow$image),
where each query has its own paired items as the relevant ones.
For visual document retrieval, following~\cite{byun2026mkp}, every adapter is trained on the ColPali training
set~\cite{faysse2025colpali} (118K query--page pairs, each page as one image; \thn samples 100K of them),
and we evaluate text query$\rightarrow$page retrieval on the ten ViDoRe-V1 sets~\cite{faysse2025colpali} and the four
English ViDoRe-V2 sets~\cite{mace2025vidorev2}, using the VLM2Vec splits~\cite{jiang2025vlm2vec}.

\nbf{Protocol}
The gallery is migrated in a random order (non-random orders are evaluated in the appendix and exhibit similar trends) and evaluated at $t\in\{0,0.1,\dots,1\}$.
At each $t$, Recall@1, $R(t)$, is the fraction of queries whose top-1 item in the merged ranking is relevant.
We report $R(0)$ of the backward system alone, $R(1)$ of the new system alone, and the AUC, the area under $R(t)$ over $t$ (trapezoidal rule), which summarizes retrieval quality over the whole migration.
Methods that learn $\rho$ use the transformed new embeddings at $t=1$, whereas the dashed new-model baseline uses the original frozen embeddings.
For image-text retrieval, all metrics are averaged over the three test sets and both retrieval directions (per-dataset results are provided in the appendix).
On ViDoRe we report nDCG@5, the benchmark's main metric, which credits relevant pages in the top five with a logarithmic discount by rank, in place of Recall@1.
 Queries arrive one at a time in random order.

\nbf{Implementation}
The adapter is a layer normalization followed by a two-layer multi-layer perceptron (MLP) of width $d_{\text{new}}$ (identity-initialized first
layer, GELU, dropout $0.1$) and an output projection to $d_{\text{old}}$, trained as in~\cite{byun2026mkp} for
10 epochs.
\thn training starts from each trained backward adapter and runs for 3,000 steps with batches of 4,096 pairs, Adam with learning
rate $10^{-4}$, $T=0.05$, $\lambda=1/30$ and a bank of $N_{\text{bank}}=16{,}384$ negatives drawn anew each step, applying
\cref{eq:merge} in both directions with a fresh migration fraction $t\sim\mathcal{U}(0.1,0.9)$ and mask per step; $\rho$ is trained
jointly. \tmc (\cref{sec:calibration}) uses $k=20$, $k'=50$ and $\beta=0.75$ for every upgrade, backward-compatible adapter and test set. Hyperparameter sensitivity analysis is in the appendix, showing that performance is robust to these choices.

\subsection{Baselines}
\begin{figure*}[t]
  \centering
  \includegraphics[width=\linewidth]{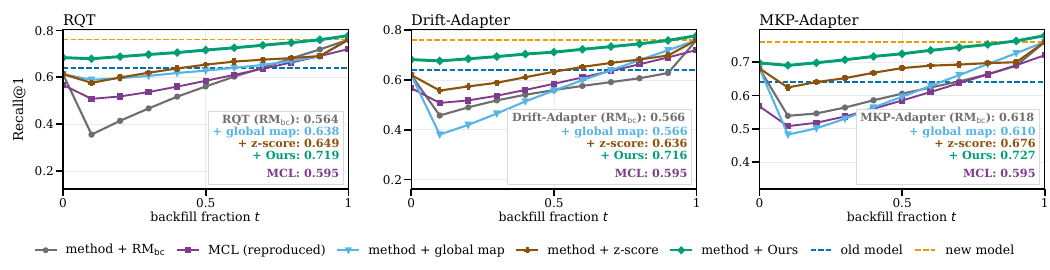}
  \caption{\textbf{Each backward-compatible adapter with plain rank merging, score normalization and ours.} Recall@1 over the backfill,
  mean over the four upgrades (COCO, Flickr30K and Urban1k, both directions). Same runs as \cref{tab:main}; lines as in
  \cref{fig:calibration_curves}; box: the Mean column of \cref{tab:main}; MCL is the same in every panel. Results per test set are given in the appendix}
  \label{fig:main_curves}
\end{figure*}
\label{sec:baselines}

Each baseline pairs a backward adapter $h$ with plain rank merging (\rmnaiveh) or rank merging augmented with score normalization.
As existing normalization methods have not been studied for mixed-database rank merging, we adapt them to construct stronger, fairer baselines.
We also compare against a reproduction of MCL, the most closely related prior work.
All methods share the same encoders and training data.

\nbf{Backward adapters} \textit{RQT.} The reverse query transform~\cite{seo2025mcl} trains $h$ to map new query embeddings onto the old ones; we train it
contrastively, with the in-batch old embeddings as negatives (with $\tau=0.07$),
\begin{equation}
  \mathcal{L}_{\text{RQT}}=-\frac{1}{B}\sum_{i=1}^{B}\log
  \frac{\exp\!\big(\cos(h(\mathbf{x}_i^{\text{new}}),\mathbf{x}_i^{\text{old}})/\tau\big)}
       {\sum_{j=1}^{B}\exp\!\big(\cos(h(\mathbf{x}_i^{\text{new}}),\mathbf{x}_j^{\text{old}})/\tau\big)}.
\end{equation}

\noindent\textit{Drift-Adapter.} Drift-Adapter~\cite{vejendla2025drift} maps new queries into the old space with a residual MLP and a diagonal
scaling, trained with the mean-squared errors.
Since our old and new dimensions differ, its identity skip connection is replaced by a linear map initialized
with the least-squares solution; all other settings follow the original.

\noindent\textit{MKP-Adapter.} MKP-Adapter~\cite{byun2026mkp} trains $h$ with a point-wise alignment term,
$\mathcal{L}_{\text{base}}=\frac{1}{N}\sum_i\big(1-\cos(h(\mathbf{x}_i^{\text{new}}),\mathbf{x}_i^{\text{old}})\big)$, plus
multi-level (point-, distance- and angle-wise) preservation terms that keep the new space's structure.

\nbf{Score normalization}
We additionally evaluate two score normalization methods applied on top of \rmnaiveh.
For the \emph{global map} (\cref{sec:calib_motivation}), we compare linear regression and moment matching (using score means and standard deviations), and report the better-performing method, moment matching.
The map is fitted once per retrieval direction using a \emph{reference set} from the adapter's training corpus, scored by both systems.
\emph{Z-score normalization}, as used in Metasearch~\cite{montague2001normalization}, standardizes each partition's scores per query and is analogous to applying only the query-side normalization of AS-norm~\cite{matejka2017asnorm}.
Additional baselines are evaluated in the appendix

\nbf{MCL}
MCL~\cite{seo2025mcl} is the most representative baseline as the prior work most directly addressing rank merging over mixed databases.
We follow its original training recipe (\cref{sec:mcl}), optimizing $h$ with \cref{eq:mcl} and jointly training $\rho$ from scratch.
Without class labels, each image--caption pair forms its own class ($\tau=0.05$; see the appendix).

\subsection{Main results}

\begin{table}[t]
\centering
\small
\setlength{\tabcolsep}{2.5pt}
\resizebox{\linewidth}{!}{%
\begin{tabular}{l c@{\hspace{8pt}} ccccc}
\toprule
& & \multicolumn{5}{c}{Image--text (COCO, Flickr30K, Urban1k)} \\
\cmidrule(l){3-7}
Backward-compatible adapter & \shortstack{ViDoRe\\V1/V2} & \shortstack{CLIP-L\\$\rightarrow$UniME} & \shortstack{E5-V\\$\rightarrow$UniME} & \shortstack{CLIP-B/16\\$\rightarrow$CLIP-L} & \shortstack{CLIP-L\\$\rightarrow$GME} & Mean \\
\midrule
RQT~\cite{seo2025mcl} & 0.477 & 0.527 & 0.745 & 0.455 & 0.530 & 0.564 \\
\quad + global map & 0.481 & 0.678 & 0.770 & 0.453 & 0.649 & 0.638 \\
\quad + z-score & 0.521 & 0.674 & 0.722 & 0.536 & 0.663 & 0.649 \\
\rowcolor{oursgray} \quad + Ours & \textbf{0.558}{\scriptsize\,\textcolor{gaincolor}{$+0.081$}} & \textbf{0.745} & \textbf{0.789} & \textbf{0.629} & \textbf{0.711} & \textbf{0.719}{\scriptsize\,\textcolor{gaincolor}{$+0.154$}} \\
\midrule
Drift-Adapter~\cite{vejendla2025drift} & 0.420 & 0.544 & 0.731 & 0.410 & 0.578 & 0.566 \\
\quad + global map & 0.426 & 0.607 & 0.620 & 0.520 & 0.518 & 0.566 \\
\quad + z-score & 0.472 & 0.658 & 0.708 & 0.552 & 0.624 & 0.636 \\
\rowcolor{oursgray} \quad + Ours & \textbf{0.535}{\scriptsize\,\textcolor{gaincolor}{$+0.115$}} & \textbf{0.745} & \textbf{0.789} & \textbf{0.628} & \textbf{0.703} & \textbf{0.716}{\scriptsize\,\textcolor{gaincolor}{$+0.150$}} \\
\midrule
MKP-Adapter~\cite{byun2026mkp} & 0.552 & 0.579 & 0.783 & 0.500 & 0.608 & 0.618 \\
\quad + global map & 0.497 & 0.623 & 0.708 & 0.584 & 0.527 & 0.610 \\
\quad + z-score & 0.544 & 0.702 & 0.745 & 0.577 & 0.681 & 0.676 \\
\rowcolor{oursgray} \quad + Ours & \textbf{0.565}{\scriptsize\,\textcolor{gaincolor}{$+0.013$}} & \textbf{0.757} & \textbf{0.797} & \textbf{0.630} & \textbf{0.724} & \textbf{0.727}{\scriptsize\,\textcolor{gaincolor}{$+0.109$}} \\
\midrule\midrule
MCL~\cite{seo2025mcl} & 0.477 & 0.612 & 0.709 & 0.494 & 0.565 & 0.595 \\
\bottomrule
\end{tabular}}
\caption{\textbf{Embedding compatibility with a mixed database across model upgrades.} Area under the nDCG@5 curve
over the backfill on ViDoRe for CLIP-L$\rightarrow$GME (mean of V1 and V2) and under the Recall@1 curve on COCO, Flickr30K and Urban1k (both
directions; per test set in the appendix). Each backward-compatible adapter is merged by plain rank merging (\rmnaiveh),
with the global map or with z-score normalization~\cite{montague2001normalization} (\cref{sec:baselines}); + Ours trains it with \thn and merges with
\tmc. Green: gain of + Ours over the plain rank merge. Best per method in bold.}
\label{tab:main}
\end{table}

\Cref{fig:calibration_curves,fig:main_curves} show the Recall@1 curves underlying the image--text AUCs in \cref{tab:main}, grouped by model upgrade (averaged over backward-compatible adapters) and by adapter (averaged over upgrades).
With plain rank merging~\cite{seo2025mcl} (\rmnaiveh, gray), all backward-compatible adapters suffer a sharp drop as migration begins.
MCL's jointly trained model (MCL, purple) shows a similar decline, with both remaining below the old model on average until at least 70\% of the gallery is migrated (\cref{fig:main_curves}).
The score normalization baselines added to \rmnaiveh provide only limited improvements.
Across upgrades (\cref{fig:calibration_curves}), the global map underperforms \rmnaiveh on E5-V$\rightarrow$UniME and CLIP-L$\rightarrow$GME, while even z-score normalization~\cite{montague2001normalization}, the strongest baseline, remains below the old model for nearly the entire backfill on E5-V$\rightarrow$UniME and CLIP-B/16$\rightarrow$CLIP-L.
Across adapters (\cref{fig:main_curves}), the global map fails to improve Drift-Adapter and MKP-Adapter, while z-score normalization surpasses the old model only after roughly half the gallery is migrated for RQT and Drift-Adapter.
Thus, these baselines remain impractical, often retrieving worse than the old model during migration.

In contrast, + Ours consistently achieves higher AUC than all baselines across all upgrades and adapters, substantially outperforming both score normalization baselines and MCL.
Notably, on average it is the only method that achieves gains from the onset of migration, with the largest improvements for adapters that perform worst under plain rank merging.
Most of these gains come from \tmc alone, without additional training (\cref{sec:ablation}).
On ViDoRe~\cite{faysse2025colpali,mace2025vidorev2} with CLIP-L$\rightarrow$GME (\cref{tab:main}, left), + Ours again achieves the best results for every adapter, outperforming MCL.
The appendix further evaluates larger galleries, R@5 and R@10, and non-random migration orders.
We also note that the AUC standard deviation is at most \sdUpgMax, across five random seeds for migration and query orders (see the appendix).

\subsection{Ablation studies}
\label{sec:ablation}

\begin{table}[t]
\centering
\small
\setlength{\tabcolsep}{2.5pt}
\resizebox{\linewidth}{!}{%
\begin{tabular}{cc cccc cc}
\toprule
\tmc & \thn & \shortstack{CLIP-L\\$\rightarrow$UniME} & \shortstack{E5-V\\$\rightarrow$UniME} & \shortstack{CLIP-B/16\\$\rightarrow$CLIP-L} & \shortstack{CLIP-L\\$\rightarrow$GME} & Mean & $\Delta$ \\
\midrule
\rowcolor{oursgray} \cmark & \cmark & \textbf{0.749} & \textbf{0.792} & \textbf{0.629} & \textbf{0.713} & \textbf{0.721} &  \\
\cmark & \xmark & 0.718 & 0.776 & 0.563 & 0.681 & 0.685 & $-0.036$ \\
\xmark & \cmark & 0.671 & 0.758 & 0.621 & 0.567 & 0.654 & $-0.067$ \\
\xmark & \xmark & 0.636 & 0.699 & 0.519 & 0.565 & 0.605 & $-0.116$ \\
\bottomrule
\end{tabular}}
\caption{\textbf{Main ablation.} Same setting as \cref{tab:main}; each adapter in the appendix \xmark{} in \tmc: the
global map, fitted once on the reference set, in its place. $\Delta$ relative to Ours.}
\label{tab:main_ablation}
\end{table}
\begin{table}[t]
\centering
\small
\setlength{\tabcolsep}{2pt}
\resizebox{\linewidth}{!}{%
\begin{tabular}{ccc cccc cc}
\toprule
(i) & (ii) & (iii) & \shortstack{CLIP-L\\$\rightarrow$UniME} & \shortstack{E5-V\\$\rightarrow$UniME} & \shortstack{CLIP-B/16\\$\rightarrow$CLIP-L} & \shortstack{CLIP-L\\$\rightarrow$GME} & Mean & $\Delta$ \\
\midrule
\rowcolor{oursgray} \cmark & \cmark & \cmark & \textbf{0.718} & \textbf{0.776} & \textbf{0.563} & \textbf{0.681} & \textbf{0.685} &  \\
\xmark & \cmark & \cmark & 0.665 & 0.723 & 0.553 & 0.635 & 0.644 & $-0.041$ \\
\cmark & \xmark & \cmark & 0.682 & 0.740 & 0.528 & 0.650 & 0.650 & $-0.035$ \\
\cmark & \cmark & \xmark & 0.700 & 0.771 & 0.553 & 0.659 & 0.671 & $-0.014$ \\
\bottomrule
\end{tabular}}
\caption{\textbf{Ablation of \tmc}. One part removed at a time (top row: \tmc only). Same setting as \cref{tab:main};
each adapter in the appendix \xmark{} in (i) tail-focused alignment: whole distributions; in (ii) partition-aware
depth: the same depth in both partitions, the best of $k=2$ to $500$ (see the appendix); in (iii) query-adaptive
intercept: $\beta=0$.}

\label{tab:tt_loo}
\end{table}
\begin{table}[t]
\centering
\footnotesize
\setlength{\tabcolsep}{4pt}
\begin{tabular}{l ccc c}
\toprule
Method & AUC & $R(0)$ & $R(1)$ & $\Delta$ \\
\midrule
\rowcolor{oursgray} Ours & \textbf{0.721} & \textbf{0.688} & \textbf{0.778} &  \\
own objective & 0.687 & 0.649 & 0.755 & $-0.033$ \\
$\mathcal{L}_{\text{MCL}}$ & 0.689 & 0.647 & 0.755 & $-0.031$ \\
\midrule
\tmc only & 0.685 & 0.638 & 0.760 & $-0.036$ \\
\bottomrule
\end{tabular}
\caption{\textbf{Ablation of \thn.} Same setting as \cref{tab:main}, merged with \tmc. Additional variants are in the appendix Rows differ only in the fine-tuning objective; $\mathcal{L}_{\text{THN}}$ (Ours), each
backward adapter's own objective (MKP-Adapter's loss, RQT's InfoNCE, Drift-Adapter's MSE) or MCL's original $\mathcal{L}_{\text{MCL}}$. \tmc only: no fine-tuning. }

\label{tab:ablation}
\end{table}

All ablations use the image--text setting of \cref{tab:main}, averaged over the three backward-compatible adapters.

\nbf{Main ablation}
\Cref{tab:main_ablation} evaluates the contribution of each component of \ours by removing components one at a time.
When \tmc is disabled (\xmark), we instead use the simplest calibration alternative, the global map (\cref{sec:calib_motivation}).
Starting from the full \ours (first row), removing \thn (second row) or replacing \tmc with the global map (third row) reduces the mean AUC, with a larger drop in the latter case.
Using the global map alone (last row) performs substantially worse than \ours, barely above plain rank merging (\cref{fig:calibration_curves}).

\nbf{\tmc}
\Cref{tab:tt_loo} presents an ablation study of the individual components of \tmc, showing that each contributes to its performance, with tail matching (i) and partition-aware depths (ii) having the largest impact.
Starting from \tmc alone (first row, without \thn), we remove one component at a time.
Removing (i), \ie matching entire distributions rather than their tails (second row), or (ii), \ie using the same depth for both partitions rather than depths matched to their sizes (third row, the best equal depth among $k=2$ to $500$; see the appendix), causes the largest performance drops across all upgrades.
This is because aligning the bulk does not necessarily align the tails that determine the merge, while equal depths correspond to different quantiles across the two partitions.
Removing (iii), the per-query intercept (fourth row), has the smallest impact.
In the appendix, we further compare \tmc alone with alternative score normalization and rank fusion techniques, including RRF~\cite{cormack2009rrf} and isotonic regression~\cite{zadrozny2002transforming}, and evaluate non-random migration orders.
\tmc clearly and consistently outperforms all baselines and maintains its advantage under non-random migration orders.

\nbf{\thn}
\Cref{tab:ablation} demonstrates the importance of our proposed tail-focused objective $\mathcal{L}_{\text{THN}}$, showing that its gains cannot be achieved by simply fine-tuning with existing objectives.
Starting from \tmc alone, we compare fine-tuning objectives for the backward adapter.
These include $\mathcal{L}_{\text{THN}}$ (\ours, \cref{eq:total}), each adapter's original objective (MKP-Adapter's loss, RQT's InfoNCE, or Drift-Adapter's MSE), and MCL's $\mathcal{L}_{\text{MCL}}$ (\cref{eq:mcl}), all using the same training budget, anchored $\rho$, and image--caption pairs from the same training corpus as the backward adapters, without any additional data.
While the original objectives and $\mathcal{L}_{\text{MCL}}$ perform comparably to \tmc alone, only $\mathcal{L}_{\text{THN}}$ achieves substantial gains.
Further ablations are in the appendix

\subsection{Computational efficiency}
\ours incurs minimal computational overhead through its two components.
\tmc requires neither training nor reference data, using only the top-$k$ scores and running statistics (counts, sums, and squared sums) from each partition per query, with cost independent of gallery size.
\thn requires only one-time training on pre-extracted embeddings (6--42 minutes on a single RTX A5000), adding at inference only a single $d_{\text{new}}\times d_{\text{new}}$ linear map $\rho$ (at most 16.8M parameters).

\section{Conclusion}

Our findings identify similarity score mismatch as a key challenge in mixed-database retrieval across heterogeneous models and modalities.
We propose \ours, which aligns the high-score tails governing merge decisions through inference-time calibration and lightweight hard-negative adapter refinement.
With minimal overhead, \ours is the only evaluated approach to outperform the old model on average throughout migration across diverse retrieval tasks, model upgrades, and adapter-based BCT methods, enabling incremental upgrades without sacrificing retrieval quality.

{\small
\bibliographystyle{ieeenat_fullname}
\bibliography{11_references}
}

\clearpage
\twocolumn[{\centering
  {\Large\bf Appendix\par}
  \vspace{1.5em}}]
\appendix
\makeatletter
\def\appfloatbarrier{\par\begingroup
  \global\@tempswafalse
  \def\@elt##1{\ifdim\wd##1>\columnwidth\else\global\@tempswatrue\fi}\@deferlist
  \if@tempswa
    \ifx\@fltovf\relax
      \if@firstcolumn\else\null\newpage\appfloatbarrier\fi
    \else\newpage\let\@fltovf\relax\appfloatbarrier\fi
  \fi\endgroup}
\makeatother
\let\origappsection\section
\renewcommand{\section}{\appfloatbarrier\origappsection}
\makeatletter
\renewcommand{\table}[1][]{\@float{table}[!htbp]}
\renewcommand{\figure}[1][]{\@float{figure}[!htbp]}
\makeatother
\section{Main results per test set}

\Cref{tab:urban1k} breaks down \cref{tab:main} by test set: COCO, Flickr30K, and Urban1k. Unlike COCO, Flickr30K, and LLaVA-LCS-558K (training corpus), Urban1k contains long, detailed captions. The results follow the same trend as \cref{tab:main}: across nearly all backward systems and test sets, our method achieves substantially larger gains over plain rank merging than the global map or z-score normalization, outperforming all baselines on every upgrade except one. The exception is E5-V$\rightarrow$UniME on Urban1k with MKP-Adapter, where the old encoder already handles long captions well and plain rank merging approaches the performance of the new model alone. Our method performs slightly worse in this case.

\makeatletter
\begingroup\renewcommand{\table}[1][]{\@float{table}[t]}\begin{table}[t]
\centering
\small
\setlength{\tabcolsep}{2.5pt}
\resizebox{\linewidth}{!}{%
\begin{tabular}{l ccccc c}
\toprule
Backward system & \shortstack{CLIP-L\\$\rightarrow$UniME} & \shortstack{E5-V\\$\rightarrow$UniME} & \shortstack{CLIP-B/16\\$\rightarrow$CLIP-L} & \shortstack{CLIP-L\\$\rightarrow$GME} & Mean & Gain \\
\midrule
\multicolumn{7}{l}{\emph{COCO}} \\
\midrule
RQT~\cite{seo2025mcl} (\rmnaiveh) & 0.440 & 0.560 & 0.350 & 0.445 & 0.449 &  \\
\quad + global map & 0.492 & 0.576 & 0.344 & 0.500 & 0.478 & \textcolor{gaincolor}{$+0.029$} \\
\quad + z-score~\cite{montague2001normalization} & 0.476 & 0.508 & 0.371 & 0.483 & 0.459 & \textcolor{gaincolor}{$+0.011$} \\
\rowcolor{oursgray} \quad + Ours & \textbf{0.578} & \textbf{0.610} & \textbf{0.477} & \textbf{0.567} & \textbf{0.558} & \textcolor{gaincolor}{$+0.109$} \\
\midrule
Drift-Adapter~\cite{vejendla2025drift} (\rmnaiveh) & 0.443 & 0.522 & 0.306 & 0.466 & 0.434 &  \\
\quad + global map & 0.463 & 0.497 & 0.387 & 0.443 & 0.448 & \textcolor{gaincolor}{$+0.013$} \\
\quad + z-score~\cite{montague2001normalization} & 0.489 & 0.500 & 0.393 & 0.481 & 0.466 & \textcolor{gaincolor}{$+0.031$} \\
\rowcolor{oursgray} \quad + Ours & \textbf{0.579} & \textbf{0.611} & \textbf{0.476} & \textbf{0.558} & \textbf{0.556} & \textcolor{gaincolor}{$+0.122$} \\
\midrule
MKP-Adapter~\cite{byun2026mkp} (\rmnaiveh) & 0.461 & 0.574 & 0.376 & 0.476 & 0.472 &  \\
\quad + global map & 0.480 & 0.549 & 0.433 & 0.448 & 0.477 & \textcolor{gaincolor}{$+0.005$} \\
\quad + z-score~\cite{montague2001normalization} & 0.512 & 0.516 & 0.411 & 0.507 & 0.487 & \textcolor{gaincolor}{$+0.015$} \\
\rowcolor{oursgray} \quad + Ours & \textbf{0.590} & \textbf{0.616} & \textbf{0.479} & \textbf{0.575} & \textbf{0.565} & \textcolor{gaincolor}{$+0.093$} \\
\midrule\midrule
MCL~\cite{seo2025mcl} & 0.480 & 0.534 & 0.373 & 0.456 & 0.461 &  \\
\midrule
\multicolumn{7}{l}{\emph{Flickr30K}} \\
\midrule
RQT~\cite{seo2025mcl} (\rmnaiveh) & 0.624 & 0.808 & 0.569 & 0.621 & 0.656 &  \\
\quad + global map & 0.755 & 0.850 & 0.588 & 0.761 & 0.739 & \textcolor{gaincolor}{$+0.083$} \\
\quad + z-score~\cite{montague2001normalization} & 0.790 & 0.839 & 0.672 & 0.785 & 0.771 & \textcolor{gaincolor}{$+0.116$} \\
\rowcolor{oursgray} \quad + Ours & \textbf{0.842} & \textbf{0.862} & \textbf{0.757} & \textbf{0.807} & \textbf{0.817} & \textcolor{gaincolor}{$+0.162$} \\
\midrule
Drift-Adapter~\cite{vejendla2025drift} (\rmnaiveh) & 0.641 & 0.842 & 0.538 & 0.703 & 0.681 &  \\
\quad + global map & 0.724 & 0.751 & 0.679 & 0.616 & 0.692 & \textcolor{gaincolor}{$+0.011$} \\
\quad + z-score~\cite{montague2001normalization} & 0.783 & 0.837 & 0.702 & 0.749 & 0.768 & \textcolor{gaincolor}{$+0.087$} \\
\rowcolor{oursgray} \quad + Ours & \textbf{0.842} & \textbf{0.861} & \textbf{0.759} & \textbf{0.798} & \textbf{0.815} & \textcolor{gaincolor}{$+0.134$} \\
\midrule
MKP-Adapter~\cite{byun2026mkp} (\rmnaiveh) & 0.701 & 0.860 & 0.653 & 0.726 & 0.735 &  \\
\quad + global map & 0.746 & 0.819 & 0.729 & 0.624 & 0.729 & \textcolor{losscolor}{$-0.005$} \\
\quad + z-score~\cite{montague2001normalization} & 0.825 & 0.850 & 0.727 & 0.799 & 0.800 & \textcolor{gaincolor}{$+0.066$} \\
\rowcolor{oursgray} \quad + Ours & \textbf{0.850} & \textbf{0.867} & \textbf{0.763} & \textbf{0.816} & \textbf{0.824} & \textcolor{gaincolor}{$+0.089$} \\
\midrule\midrule
MCL~\cite{seo2025mcl} & 0.753 & 0.809 & 0.637 & 0.682 & 0.720 &  \\
\midrule
\multicolumn{7}{l}{\emph{Urban1k}} \\
\midrule
RQT~\cite{seo2025mcl} (\rmnaiveh) & 0.518 & 0.867 & 0.447 & 0.523 & 0.589 &  \\
\quad + global map & 0.788 & 0.884 & 0.427 & 0.686 & 0.696 & \textcolor{gaincolor}{$+0.108$} \\
\quad + z-score~\cite{montague2001normalization} & 0.756 & 0.817 & 0.565 & 0.723 & 0.715 & \textcolor{gaincolor}{$+0.127$} \\
\rowcolor{oursgray} \quad + Ours & \textbf{0.815} & \textbf{0.894} & \textbf{0.654} & \textbf{0.760} & \textbf{0.781} & \textcolor{gaincolor}{$+0.192$} \\
\midrule
Drift-Adapter~\cite{vejendla2025drift} (\rmnaiveh) & 0.548 & 0.830 & 0.386 & 0.564 & 0.582 &  \\
\quad + global map & 0.634 & 0.613 & 0.494 & 0.495 & 0.559 & \textcolor{losscolor}{$-0.023$} \\
\quad + z-score~\cite{montague2001normalization} & 0.703 & 0.786 & 0.560 & 0.643 & 0.673 & \textcolor{gaincolor}{$+0.091$} \\
\rowcolor{oursgray} \quad + Ours & \textbf{0.814} & \textbf{0.896} & \textbf{0.647} & \textbf{0.752} & \textbf{0.777} & \textcolor{gaincolor}{$+0.196$} \\
\midrule
MKP-Adapter~\cite{byun2026mkp} (\rmnaiveh) & 0.577 & \textbf{0.914} & 0.472 & 0.622 & 0.646 &  \\
\quad + global map & 0.642 & 0.755 & 0.590 & 0.510 & 0.624 & \textcolor{losscolor}{$-0.022$} \\
\quad + z-score~\cite{montague2001normalization} & 0.771 & 0.867 & 0.593 & 0.737 & 0.742 & \textcolor{gaincolor}{$+0.096$} \\
\rowcolor{oursgray} \quad + Ours & \textbf{0.830} & 0.907 & \textbf{0.649} & \textbf{0.780} & \textbf{0.792} & \textcolor{gaincolor}{$+0.145$} \\
\midrule\midrule
MCL~\cite{seo2025mcl} & 0.603 & 0.784 & 0.472 & 0.557 & 0.604 &  \\
\bottomrule
\end{tabular}}
\caption{\textbf{Main results per test set.} \Cref{tab:main} (Recall@1 AUC over the backfill, both directions)
on each test set separately: COCO, Flickr30K and Urban1k; \cref{tab:main}
is their mean. Each backward method uses plain rank merging (\rmnaiveh) in its first row. + Ours: our method,
\ours. Gain: mean AUC minus that of \rmnaiveh of the same method. Best per method in bold.}
\label{tab:urban1k}
\end{table}\endgroup
\makeatother

\begin{figure*}[t]
  \centering
  \includegraphics[width=\linewidth]{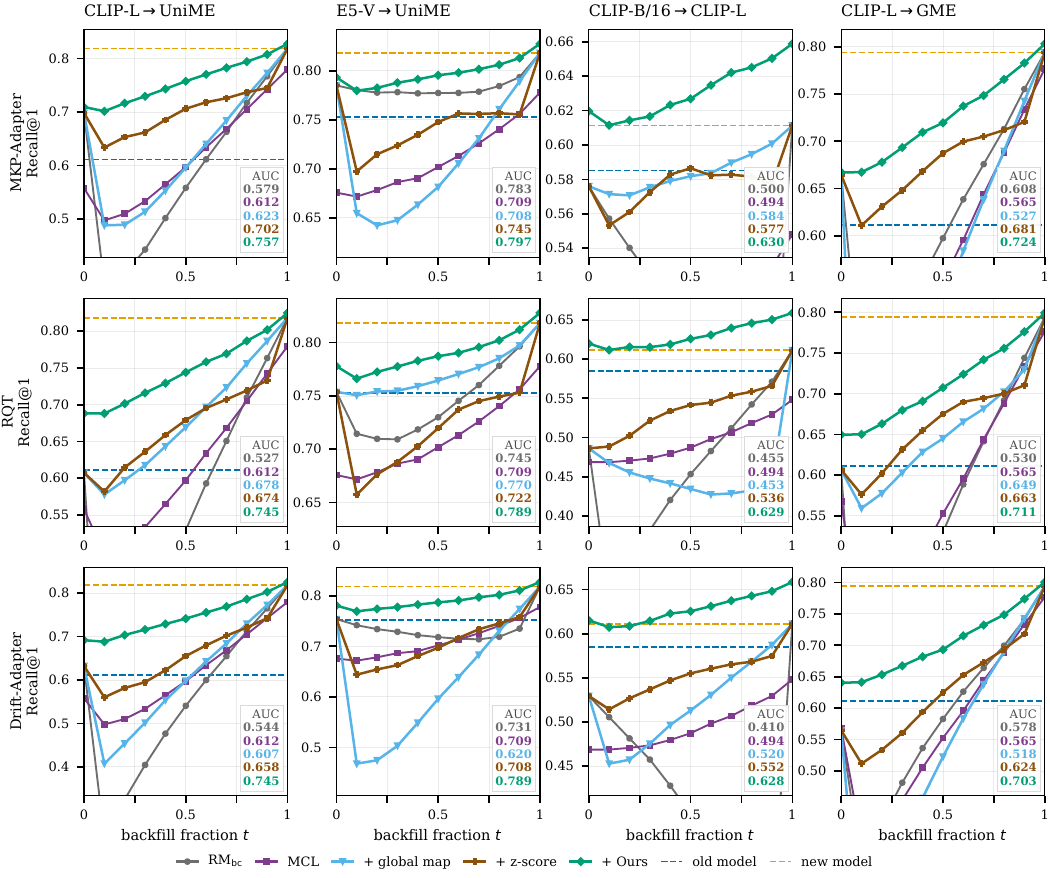}
  \caption{\textbf{Calibration curves averaged over COCO, Flickr30K and Urban1k} (both directions, the six test sets
  and directions weighted equally, as in \cref{tab:main}), for MKP-Adapter (top), RQT (middle)
  and Drift-Adapter (bottom) as the backward system, as in \cref{fig:calibration_curves}; the global map is fitted on
  LLaVA-LCS-558K. Dashed: old and new model alone; box: AUC.}
  \label{fig:appendix_all3_curves}
\end{figure*}

\Cref{fig:appendix_all3_curves} shows performance throughout backfilling for each backward system. It follows the findings of the manuscript: only our method consistently performs well across backward adapters and upgrades, and \ours achieves the highest AUC in every panel. In contrast, each baseline normalization fails in certain settings. Z-score normalization underperforms plain rank merging on E5-V$\rightarrow$UniME for nearly every backward system and test set. On CLIP-B/16$\rightarrow$CLIP-L and CLIP-L$\rightarrow$GME, z-score normalization helps only late in the backfill. Right after migration starts ($t=0.1$), it falls to or below the old model alone for every backward system. On CLIP-L$\rightarrow$GME it rises above the old model only from $t=0.2$--$0.5$, depending on the backward system, and on CLIP-B/16$\rightarrow$CLIP-L it stays at or below the old model through the whole backfill. Near the end it levels off again: at $t=0.9$ on CLIP-L$\rightarrow$GME, it is below plain rank merging for every backward system. Both ends are where one partition is small. Z-score normalization standardizes each partition regardless of its size, so it likely removes the partition-size effect that \tmc keeps with its partition-aware depths (\cref{sec:method}). The global map similarly underperforms on E5-V$\rightarrow$UniME and CLIP-L$\rightarrow$GME with MKP-Adapter and Drift-Adapter across all test sets, with the largest degradation on Urban1k.

\section{Ablations for every backward system}

\begin{table}[t]
\centering
\small
\setlength{\tabcolsep}{2.5pt}
\resizebox{\linewidth}{!}{%
\begin{tabular}{cc cccc cc}
\toprule
\tmc & \thn & \shortstack{CLIP-L\\$\rightarrow$UniME} & \shortstack{E5-V\\$\rightarrow$UniME} & \shortstack{CLIP-B/16\\$\rightarrow$CLIP-L} & \shortstack{CLIP-L\\$\rightarrow$GME} & Mean & $\Delta$ \\
\midrule
\multicolumn{8}{l}{\textbf{Backward system: MKP-Adapter}} \\
\rowcolor{oursgray} \cmark & \cmark & \textbf{0.757} & \textbf{0.797} & \textbf{0.630} & \textbf{0.724} & \textbf{0.727} &  \\
\cmark & \xmark & 0.746 & 0.789 & 0.587 & 0.717 & 0.710 & $-0.017$ \\
\xmark & \cmark & 0.678 & 0.773 & 0.630 & 0.560 & 0.660 & $-0.067$ \\
\xmark & \xmark & 0.623 & 0.708 & 0.584 & 0.527 & 0.610 & $-0.117$ \\
\midrule
\multicolumn{8}{l}{\textbf{Backward system: RQT}} \\
\rowcolor{oursgray} \cmark & \cmark & \textbf{0.745} & \textbf{0.789} & \textbf{0.629} & \textbf{0.711} & \textbf{0.719} &  \\
\cmark & \xmark & 0.702 & 0.770 & 0.541 & 0.681 & 0.673 & $-0.045$ \\
\xmark & \cmark & 0.676 & 0.768 & 0.622 & 0.595 & 0.665 & $-0.054$ \\
\xmark & \xmark & 0.678 & 0.770 & 0.453 & 0.649 & 0.638 & $-0.081$ \\
\midrule
\multicolumn{8}{l}{\textbf{Backward system: Drift-Adapter}} \\
\rowcolor{oursgray} \cmark & \cmark & \textbf{0.745} & \textbf{0.789} & \textbf{0.628} & \textbf{0.703} & \textbf{0.716} &  \\
\cmark & \xmark & 0.706 & 0.770 & 0.562 & 0.645 & 0.671 & $-0.046$ \\
\xmark & \cmark & 0.658 & 0.731 & 0.610 & 0.547 & 0.637 & $-0.079$ \\
\xmark & \xmark & 0.607 & 0.620 & 0.520 & 0.518 & 0.566 & $-0.150$ \\
\bottomrule
\end{tabular}}
\caption{\textbf{Main ablation for every backward system} (\cref{tab:main_ablation}; Recall@1 AUC, $\Delta$ relative
to Ours, gray; \tmc and \thn are ours). \thn trains each backward system's adapter with \thn; \tmc \xmark: the global
map, one affine map fitted once on LLaVA-LCS-558K (no test data), in place of \tmc.}
\label{tab:abl_bk_main}
\end{table}
\begin{table}[t]
\centering
\small
\setlength{\tabcolsep}{2.5pt}
\resizebox{\linewidth}{!}{%
\begin{tabular}{ccc cccc cc}
\toprule
(i) & (ii) & (iii) & \shortstack{CLIP-L\\$\rightarrow$UniME} & \shortstack{E5-V\\$\rightarrow$UniME} & \shortstack{CLIP-B/16\\$\rightarrow$CLIP-L} & \shortstack{CLIP-L\\$\rightarrow$GME} & Mean & $\Delta$ \\
\midrule
\multicolumn{9}{l}{\textbf{Backward system: MKP-Adapter}} \\
\rowcolor{oursgray} \cmark & \cmark & \cmark & \textbf{0.746} & \textbf{0.789} & \textbf{0.587} & \textbf{0.717} & \textbf{0.710} &  \\
\xmark & \cmark & \cmark & 0.702 & 0.753 & 0.586 & 0.682 & 0.681 & $-0.029$ \\
\cmark & \xmark & \cmark & 0.706 & 0.750 & 0.550 & 0.677 & 0.671 & $-0.039$ \\
\cmark & \cmark & \xmark & 0.736 & 0.785 & 0.584 & 0.701 & 0.701 & $-0.008$ \\
\midrule
\multicolumn{9}{l}{\textbf{Backward system: RQT}} \\
\rowcolor{oursgray} \cmark & \cmark & \cmark & \textbf{0.702} & \textbf{0.770} & \textbf{0.541} & \textbf{0.681} & \textbf{0.673} &  \\
\xmark & \cmark & \cmark & 0.667 & 0.719 & 0.529 & 0.650 & 0.641 & $-0.032$ \\
\cmark & \xmark & \cmark & 0.669 & 0.734 & 0.507 & 0.649 & 0.640 & $-0.034$ \\
\cmark & \cmark & \xmark & 0.685 & 0.765 & 0.527 & 0.656 & 0.658 & $-0.015$ \\
\midrule
\multicolumn{9}{l}{\textbf{Backward system: Drift-Adapter}} \\
\rowcolor{oursgray} \cmark & \cmark & \cmark & \textbf{0.706} & \textbf{0.770} & \textbf{0.562} & \textbf{0.645} & \textbf{0.671} &  \\
\xmark & \cmark & \cmark & 0.627 & 0.696 & 0.545 & 0.573 & 0.610 & $-0.061$ \\
\cmark & \xmark & \cmark & 0.672 & 0.736 & 0.525 & 0.624 & 0.639 & $-0.031$ \\
\cmark & \cmark & \xmark & 0.680 & 0.762 & 0.548 & 0.620 & 0.653 & $-0.018$ \\
\bottomrule
\end{tabular}}
\caption{\textbf{Ablation of \tmc-only (ours) for every backward system.} \Cref{tab:tt_loo} for each backward system
separately: one part removed at a time (Recall@1 AUC; $\Delta$ relative to Ours (\tmc), gray). (i)~Tail-focused
alignment, \xmark: whole distributions; (ii)~partition-aware depth, \xmark: the same depth in both partitions, the
best $k$ of a sweep over 2 to 500 for each backward system; (iii)~query-adaptive intercept, \xmark: $\beta=0$.}
\label{tab:abl_bk_tmc}
\end{table}
\begin{table}[t]
\centering
\small
\setlength{\tabcolsep}{3pt}
\resizebox{\linewidth}{!}{%
\begin{tabular}{l cccccccc}
\toprule
Tail depth $k$ & 2 & 5 & 10 & 20 & 50 & 100 & 200 & 500 \\
\midrule
Same depth & \textbf{0.650} & 0.648 & 0.640 & 0.628 & 0.610 & 0.613 & 0.623 & 0.634 \\
\midrule
\rowcolor{oursgray} \tmc & \multicolumn{8}{c}{\textbf{0.685} (default setting: $k=20$, $k'=50$)} \\
\bottomrule
\end{tabular}}
\caption{\textbf{Tail depth of \tmc.} Setting of \cref{tab:tt_loo} (Recall@1 AUC, mean over the four upgrades and the
three backward systems without \thn). Same depth: both partitions use their top-$k$, for each depth $k$
(and $k'=2.5k$ for the per-query intercept); the best ($k=2$, bold) is the one used for (ii)~\xmark{} in
\cref{tab:tt_loo}. \tmc (gray): its default setting only, $k=20$ split by partition size as
$k_n=\operatorname{round}(kt)$, $k_o=\operatorname{round}(k(1-t))$ (\cref{eq:depth}), and $k'=50$.}
\label{tab:depth_sweep}
\end{table}

\subsection{Detailed results on main ablation and \tmc}

\nbf{Main ablation}
\Cref{tab:abl_bk_main} reports the results of \cref{tab:main_ablation} for each backward system, confirming the overall trend: both \tmc and \thn consistently contribute to performance across all backward systems. Each independently improves the mean AUC, removing either leads to a substantial drop, and removing both results in the largest degradation.

\nbf{Ablation of \tmc}
\Cref{tab:abl_bk_tmc} reports the standalone ablation results of \tmc (without \thn) for each backward system, extending \cref{tab:tt_loo}. Removing any of the three components reduces AUC across all backward systems. Partition-aware depth is particularly important for MKP-Adapter and RQT, while tail-focused alignment has the largest impact on Drift-Adapter. These results demonstrate that all three components contribute to \tmc's effectiveness.

\nbf{Tail depth}
\Cref{tab:depth_sweep} varies the tail depth $k$ when both partitions use the same depth, and compares it with \tmc's partition-aware depths at its default setting ($k=20$, $k'=50$). Among the same depths, $k=2$ performs best and is therefore used for (ii)~\xmark{} in \cref{tab:tt_loo}. \tmc outperforms the same depth at every $k$.
\subsection{Ablation of \thn}

\begin{table}[t]
\centering
\small
\setlength{\tabcolsep}{2.5pt}
\resizebox{\linewidth}{!}{%
\begin{tabular}{l cccc c}
\toprule
Variant & \shortstack{CLIP-L\\$\rightarrow$UniME} & \shortstack{E5-V\\$\rightarrow$UniME} & \shortstack{CLIP-B/16\\$\rightarrow$CLIP-L} & \shortstack{CLIP-L\\$\rightarrow$GME} & Mean \\
\midrule
\multicolumn{6}{l}{\textbf{MKP-Adapter}} \\
\rowcolor{oursgray} Ours & \textbf{0.757} & \textbf{0.797} & \textbf{0.630} & \textbf{0.724} & \textbf{0.727} \\
no bank & 0.753 & 0.794 & 0.628 & 0.721 & 0.724 \\
free $\rho$ & 0.754 & 0.792 & 0.627 & 0.721 & 0.724 \\
\midrule
own loss & 0.736 & 0.786 & 0.569 & 0.705 & 0.699 \\
$\mathcal{L}_{\text{MCL}}$ & 0.723 & 0.774 & 0.567 & 0.712 & 0.694 \\
\midrule
start (no fine-tuning) & 0.746 & 0.789 & 0.587 & 0.717 & 0.710 \\
\midrule
\multicolumn{6}{l}{\textbf{RQT}} \\
\rowcolor{oursgray} Ours & \textbf{0.745} & \textbf{0.789} & \textbf{0.629} & \textbf{0.711} & \textbf{0.719} \\
no bank & 0.739 & 0.785 & 0.626 & 0.709 & 0.715 \\
free $\rho$ & 0.740 & 0.782 & 0.626 & 0.707 & 0.714 \\
\midrule
own loss & 0.713 & 0.772 & 0.564 & 0.693 & 0.686 \\
$\mathcal{L}_{\text{MCL}}$ & 0.704 & 0.766 & 0.563 & 0.697 & 0.682 \\
\midrule
start (no fine-tuning) & 0.702 & 0.770 & 0.541 & 0.681 & 0.673 \\
\midrule
\multicolumn{6}{l}{\textbf{Drift-Adapter}} \\
\rowcolor{oursgray} Ours & \textbf{0.745} & \textbf{0.789} & \textbf{0.628} & \textbf{0.703} & \textbf{0.716} \\
no bank & 0.744 & 0.789 & 0.627 & 0.700 & 0.715 \\
free $\rho$ & 0.740 & 0.785 & 0.623 & 0.699 & 0.712 \\
\midrule
own loss & 0.718 & 0.773 & 0.562 & 0.655 & 0.677 \\
$\mathcal{L}_{\text{MCL}}$ & 0.722 & 0.774 & 0.567 & 0.704 & 0.692 \\
\midrule
start (no fine-tuning) & 0.706 & 0.770 & 0.562 & 0.645 & 0.671 \\
\bottomrule
\end{tabular}}
\caption{\textbf{Ablation of \thn (ours) for every backward system}: the rows of \cref{tab:ablation} and further
variants with each backward system as the start, per upgrade (Recall@1 AUC with \tmc, COCO, Flickr30K and Urban1k in
both directions), and their mean.}
\label{tab:ablation_bk}
\end{table}

\Cref{tab:ablation_bk} breaks \cref{tab:ablation} down by backward system and upgrade, with further variants. Ours
is \ours: the backward system fine-tuned with $\mathcal{L}_{\text{THN}}$ (\cref{eq:total}) and merged with \tmc.
We first remove parts of \thn. ``No bank'' trains without the negative bank, and ``free $\rho$'' still trains $\rho$
but without its regularizer toward the identity (\cref{sec:method}). Each lowers the mean AUC slightly for every backward system.
We then replace the objective, fine-tuning each backward system with its adapter's own objective (``own loss'') or
with $\mathcal{L}_{\text{MCL}}$. Either changes the mean AUC little from the starting point without fine-tuning, or even lowers it (MKP-Adapter), whereas fine-tuning with $\mathcal{L}_{\text{THN}}$ considerably improves mean AUC and achieves the best mean AUC for every backward system.
\section{Calibration baselines vs \tmc}

\begin{table}[t]
\centering
\footnotesize
\setlength{\tabcolsep}{3pt}
\begin{tabular}{@{}ll cccc@{}}
\toprule
Calibration & Scores & RQT & MKP & Drift & Avg \\
\midrule
None (\rmnaiveh) & -- & 0.564 & 0.618 & 0.566 & 0.583 \\
\midrule
\multicolumn{6}{@{}l}{\textit{Global (LLaVA-LCS-558K, fixed for all $t$)}} \\
\hspace{0.5em}Linear regression & whole & 0.461 & 0.616 & 0.539 & 0.539 \\
\hspace{0.5em}Linear regression & top-20 & 0.529 & 0.641 & 0.571 & 0.580 \\
\hspace{0.5em}Platt scaling~\cite{platt1999probabilistic} & whole & 0.634 & 0.656 & 0.606 & 0.632 \\
\hspace{0.5em}Isotonic regression~\cite{zadrozny2002transforming} & whole & 0.640 & 0.654 & 0.609 & 0.634 \\
\hspace{0.5em}Moment matching & whole & 0.638 & 0.610 & 0.566 & 0.605 \\
\hspace{0.5em}Moment matching & top-20 & 0.619 & 0.691 & 0.624 & 0.645 \\
\midrule
\multicolumn{6}{@{}l}{\textit{Test statistics (current query only)}} \\
\hspace{0.5em}Z-score / AS-norm~\cite{montague2001normalization} & whole & 0.649 & 0.676 & 0.636 & 0.654 \\
\hspace{0.5em}Z-score / AS-norm~\cite{montague2001normalization} & top-20 & 0.606 & 0.636 & 0.609 & 0.617 \\
\hspace{0.5em}RRF~\cite{cormack2009rrf} & ranks & 0.455 & 0.480 & 0.461 & 0.465 \\
\midrule
\multicolumn{6}{@{}l}{\textit{Ours (current query + running statistics)}} \\
\rowcolor{oursgray} \hspace{0.5em}\tmc & size-scaled tails & \textbf{0.673} & \textbf{0.710} & \textbf{0.671} & \textbf{0.685} \\
\bottomrule
\end{tabular}
\caption{\textbf{Calibration baselines.} Setting of \cref{tab:main} (Recall@1 AUC, mean over the four upgrades),
backward systems without \thn (MKP: MKP-Adapter, Drift: Drift-Adapter). Moment matching maps the new system's scores to the
old system's mean and standard deviation. Z-score / AS-norm standardizes each partition's whole list (its top 1000 scores) or its top-20 (\tmc's tail depth) per query
(metasearch's zero-mean unit-variance normalization \cite{montague2001normalization}, and query-side AS-norm~\cite{matejka2017asnorm});
 since each item lies in one list, the merge compares the two top-1 z-scores. Reciprocal rank fusion (RRF~\cite{cormack2009rrf}) gives both partitions' top-1 the same value, so the merge picks either at random. Global maps use 8,000 LLaVA-LCS-558K queries against 8,000 of its items as references.
Scores: which scores the map is fitted to; Platt scaling~\cite{platt1999probabilistic} and isotonic
regression~\cite{zadrozny2002transforming} map
each system's score to P(relevant). Best per column in bold.}
\label{tab:tmc_baselines}
\end{table}
\begin{table*}[t]
\centering
\small
\setlength{\tabcolsep}{3pt}
\resizebox{\textwidth}{!}{%
\begin{tabular}{l ccccc@{\hspace{10pt}} ccccc}
\toprule
& \multicolumn{5}{c}{R@5} & \multicolumn{5}{c}{R@10} \\
\cmidrule(lr){2-6}\cmidrule(lr){7-11}
Backward system & \shortstack{CLIP-L\\$\rightarrow$UniME} & \shortstack{E5-V\\$\rightarrow$UniME} & \shortstack{CLIP-B/16\\$\rightarrow$CLIP-L} & \shortstack{CLIP-L\\$\rightarrow$GME} & Mean & \shortstack{CLIP-L\\$\rightarrow$UniME} & \shortstack{E5-V\\$\rightarrow$UniME} & \shortstack{CLIP-B/16\\$\rightarrow$CLIP-L} & \shortstack{CLIP-L\\$\rightarrow$GME} & Mean \\
\midrule
RQT~\cite{seo2025mcl} & 0.621 & 0.905 & 0.636 & 0.665 & 0.706 & 0.647 & 0.942 & 0.698 & 0.716 & 0.751 \\
\quad + global map & 0.857 & 0.917 & 0.692 & 0.839 & 0.826 & 0.909 & 0.952 & 0.771 & 0.889 & 0.880 \\
\quad + z-score~\cite{montague2001normalization} & 0.854 & 0.889 & 0.767 & 0.850 & 0.840 & 0.905 & 0.932 & 0.839 & 0.900 & 0.894 \\
\rowcolor{oursgray} \quad + Ours & \textbf{0.905} & \textbf{0.930} & \textbf{0.836} & \textbf{0.891} & \textbf{0.891} & \textbf{0.943} & \textbf{0.960} & \textbf{0.895} & \textbf{0.931} & \textbf{0.932} \\
\midrule
Drift-Adapter~\cite{vejendla2025drift} & 0.673 & 0.892 & 0.615 & 0.743 & 0.731 & 0.726 & 0.933 & 0.692 & 0.797 & 0.787 \\
\quad + global map & 0.776 & 0.823 & 0.744 & 0.639 & 0.745 & 0.834 & 0.891 & 0.819 & 0.688 & 0.808 \\
\quad + z-score~\cite{montague2001normalization} & 0.846 & 0.883 & 0.777 & 0.815 & 0.830 & 0.899 & 0.928 & 0.847 & 0.871 & 0.886 \\
\rowcolor{oursgray} \quad + Ours & \textbf{0.903} & \textbf{0.931} & \textbf{0.835} & \textbf{0.880} & \textbf{0.887} & \textbf{0.941} & \textbf{0.960} & \textbf{0.894} & \textbf{0.923} & \textbf{0.930} \\
\midrule
MKP-Adapter~\cite{byun2026mkp} & 0.747 & 0.921 & 0.724 & 0.797 & 0.797 & 0.812 & 0.953 & 0.800 & 0.857 & 0.855 \\
\quad + global map & 0.804 & 0.898 & 0.803 & 0.659 & 0.791 & 0.863 & 0.943 & 0.868 & 0.718 & 0.848 \\
\quad + z-score~\cite{montague2001normalization} & 0.876 & 0.902 & 0.795 & 0.862 & 0.859 & 0.922 & 0.940 & 0.861 & 0.909 & 0.908 \\
\rowcolor{oursgray} \quad + Ours & \textbf{0.911} & \textbf{0.933} & \textbf{0.836} & \textbf{0.893} & \textbf{0.893} & \textbf{0.946} & \textbf{0.962} & \textbf{0.895} & \textbf{0.932} & \textbf{0.934} \\
\midrule\midrule
MCL~\cite{seo2025mcl} & 0.802 & 0.888 & 0.726 & 0.755 & 0.793 & 0.862 & 0.930 & 0.808 & 0.822 & 0.855 \\
\bottomrule
\end{tabular}}
\caption{\textbf{Recall@5 and Recall@10.} Area under the R@5 and R@10 curves over the backfill in the image--text
setting of \cref{tab:main} (COCO, Flickr30K and Urban1k, both directions). Rows as in \cref{tab:main}. Best per method
in bold.}
\label{tab:recallk}
\end{table*}

\Cref{tab:tmc_baselines} compares \tmc with alternative methods for aligning the two systems' scores before merging, evaluated for each backward system without \thn (the last row is \tmc alone).
These methods differ in which scores are used to fit the transformation and when it is fitted. We discuss the results averaged over the three backward systems below.

\nbf{No calibration}
``None (\rmnaiveh)'' (first row) compares the raw cosine similarities of the two lists, the plain rank merge.

\nbf{Global (LLaVA-LCS-558K, fixed for all $t$)}
These methods fit one map per upgrade, backward system and retrieval direction only on LLaVA-LCS-558K, the adapters' training data, with no
test data: 8,000 of its queries are scored by both systems against 8,000 of its items as references (the
new system with the new embeddings, the backward system with $h(\mathbf{q}^{\text{new}})$ against the old
embeddings), and the map is then applied unchanged at every $t$. ``Whole'' fits a map to all these scores, and
``top-20'' to each query's top-20 scores.

\textit{With pair information.} The first three methods are fitted with the pairs of LLaVA-LCS-558K: linear
regression uses the two systems' scores of the same query--item pair, and Platt scaling~\cite{platt1999probabilistic} and isotonic
regression~\cite{zadrozny2002transforming} use which item is each query's pair as relevance labels.
``Linear regression'', the most basic, maps the new system's scores affinely, $as+b$, fitted by least squares
to the backward system's score of the same query--item pair, over all pairs (``whole'') or over the new system's
top-20 items of each query (``top-20'').
``Platt scaling''~\cite{platt1999probabilistic} extends it to probabilities: it passes an affine function of each
system's score through the logistic function, $\sigma(\alpha s+\gamma)$, which maps the score to the probability that
the item is relevant, and fits $\alpha$ and $\gamma$ to relevance labels rather than to the other system's scores,
minimizing the binary cross-entropy. Each query's paired item is relevant (8,000 pairs) and every other query--item
pair is irrelevant (2 million of them sampled at random, each weighted to stand for all $8{,}000\times7{,}999$); the
fit is per system and over all these scores (``whole''). ``Isotonic regression''~\cite{zadrozny2002transforming}
drops the functional form as well and assumes only that a higher score never means a lower probability of relevance.
It fits to the same weighted labels the non-decreasing function of the score closest to them in weighted least
squares, found by pool adjacent violators: going up the sorted training scores, wherever the fitted values would
decrease, adjacent scores are pooled into one block whose value is the weighted mean of their labels, the fraction of
relevant pairs in the block, until the values never decrease. The result is a non-decreasing step function, the
empirical probability of relevance of each block; at test time a score is mapped by linear interpolation between
the blocks, and clipped to the end values outside the range of the training scores. 
For both, the merge compares the probabilities of the two top-1s.

``Moment matching'' is the version that needs no pairs: it fits the affine map, $as+b$, from the two score distributions alone, matching the
mean and standard deviation of the new system's scores to those of the backward system,
$a=\operatorname{std}^{\text{old}}/\operatorname{std}^{\text{new}}$ and
$b=\operatorname{mean}^{\text{old}}-a\operatorname{mean}^{\text{new}}$ (``whole'' is the default setting, the global map of \cref{tab:main}).

Unlike \tmc, all these maps are fixed once fitted: they follow neither the test queries, nor the sizes of the
partitions, which change with $t$, nor the individual query. Except for the top-20 variants, they are also fitted
over all scores, most of them of irrelevant items, rather than at the top of each list where the merge decides; and
the first three need pair information, which \tmc does not.

\nbf{Test statistics (current query only)}
These methods are fitted at test time to the current query's two lists, with no other data.
``Z-score / AS-norm'' is the test-time version of global moment matching: rather than matching distributions of
LLaVA-LCS-558K, it standardizes each of the two separate partitions, the new system's scores over the migrated
items and the backward system's scores over the remaining ones, by the mean and standard deviation of the current
query's top-$N$ scores in that partition, $(s-\mu)/\sigma$, and compares the two top-1 z-scores.
This is the zero-mean, unit-variance normalization of metasearch~\cite{montague2001normalization}, known
as Z-normalization (Z-norm) in speaker verification~\cite{auckenthaler2000score}, and query-side AS-norm~\cite{sturim2005speaker,matejka2017asnorm}, which standardizes a score by the query's top-$N$ cohort; since each item
lies in exactly one partition, both reduce to this comparison.
``Whole'' is the default setting, the z-score normalization baseline of \cref{tab:main}: it takes $N=1000$, the same setup as in~\cite{montague2001normalization}; ``top-20'' takes $N=20$, \tmc's tail depth. ``RRF'', reciprocal rank fusion~\cite{cormack2009rrf},
replaces each score by $1/(60+r)$ with $r$ the rank in its list. 
It gives the two top-1s the same value, so the merge picks either at random (a tie counts half for each).

\tmc is fitted at test time from score statistics alone, without
training data, but it differs from them in three ways (\cref{sec:calibration}). First, rather than standardizing each partition by its own statistics, it maps one
system's scores onto the other's with a scale and an intercept. Standardization
divides each partition's scores by the mean and standard deviation of its own top-$N$
scores, but these statistics are not comparable across partitions: the more items a
partition holds, the higher its top-$N$ scores tend to be. Standardizing each partition
by its own statistics can therefore suppress the naturally higher top scores of the larger
partition. Under random migration, the larger partition is more likely to contain the
relevant item, so these higher scores reflect a genuine advantage, which \tmc preserves.
Second, it matches only the upper tails, where the merge is decided, at depths proportional to the partition sizes, $k_n=\operatorname{round}(kt)$ and $k_o=\operatorname{round}(k(1-t))$, so that both tails cover the same top fraction of their partitions. The global maps and z-score normalization, by contrast, use all scores or a fixed depth (the top-20), regardless of the partition's size or where the merge is decided.
Third, its scale is pooled over all queries served so far through running statistics, and only its intercept adapts to the current query, shrunk toward the pooled intercept by $\beta$. 

\nbf{Results}
\tmc performs best for every backward system. Among the global maps, regressing one
system's scores on the other's performs on par with or below the raw merge, while mapping
each system's scores to a probability of relevance improves on the raw merge but still
falls short of \tmc. For moment matching, matching the upper tails rather than the whole
distributions helps on average, but a fixed map adapts neither to the test queries nor to
the partitions. Among the methods fitted on test statistics, z-score normalization over
nearly the whole list is the strongest baseline on average, and restricting it to the
top-20 makes it worse: a fixed depth covers a larger fraction of the smaller partition
than of the larger one, so the two tails are not comparable. Over the course of the backfill
(\cref{fig:calibration_curves}), z-score normalization reduces the early drop of the raw
merge (in certain cases) but does not eliminate it, and its AUC remains below that of \tmc on every upgrade
(\cref{tab:main_ablation}). Across backward methods (\cref{fig:main_curves}), it usually improves over the raw merge and the global map but remains well below \tmc.
RRF falls below even the raw merge: it assigns the same score
to each partition's top item, so the top of the merged list is decided by tie-breaking
rather than by the scores or the partition sizes. With disjoint partitions, RRF in fact
reduces to interleaving the two lists.
\section{Recall@5 and Recall@10}

\Cref{tab:recallk} repeats the image--text part of \cref{tab:main} at Recall@5 and
Recall@10. The results follow the Recall@1 trend of \cref{tab:main}: at both
cutoffs, + Ours performs best in every column for every backward method and remains
above MCL.
\section{Sensitivity of the settings}

\begin{table}[t]
\centering
\small
\setlength{\tabcolsep}{3pt}
\begin{tabular}{l cc}
\toprule
Setting & Ours (\tmc) & Ours \\
\midrule
\multicolumn{3}{l}{\emph{$k$, the others at their defaults}} \\
\quad $k=5$ & 0.709 & 0.725 \\
\quad $k=10$ & 0.709 & 0.726 \\
\rowcolor{oursgray} \quad $k=20$ & 0.710 & 0.727 \\
\quad $k=50$ & 0.710 & 0.728 \\
\quad $k=100$ & 0.710 & 0.727 \\
\multicolumn{3}{l}{\emph{$k'$, the others at their defaults}} \\
\quad $k'=10$ & 0.705 & 0.723 \\
\quad $k'=20$ & 0.707 & 0.725 \\
\rowcolor{oursgray} \quad $k'=50$ & 0.710 & 0.727 \\
\quad $k'=100$ & 0.709 & 0.727 \\
\quad $k'=200$ & 0.708 & 0.725 \\
\multicolumn{3}{l}{\emph{$\beta$, the others at their defaults}} \\
\quad $\beta=0$ & 0.701 & 0.720 \\
\quad $\beta=0.25$ & 0.705 & 0.724 \\
\quad $\beta=0.5$ & 0.708 & 0.726 \\
\rowcolor{oursgray} \quad $\beta=0.75$ & 0.710 & 0.727 \\
\quad $\beta=1$ & 0.708 & 0.725 \\
\midrule
oracle: $k{=}50$, $k'{=}50$, $\beta{=}0.75$ & 0.710 & 0.728 \\
no calibration (\rmnaiveh) & 0.618 & 0.709 \\
\bottomrule
\end{tabular}
\caption{\textbf{Sensitivity of \tmc's settings.} Recall@1 AUC, mean of the four upgrades; MKP-Adapter as the backward
system (trained with \thn in the last column). Test: COCO, Flickr30K and Urban1k, both
directions. Oracle: the best setting on the test set, searched over the full grid $k\in\{5,\dots,100\}$, $k'\in\{10,\dots,200\}$,
$\beta\in\{0,\dots,1\}$. Gray: the paper's setting.}
\label{tab:hparam}
\end{table}

\tmc uses one setting, $k=20$, $k'=50$ and $\beta=0.75$, for every upgrade, backward system and test set.
\Cref{tab:hparam} changes one of them at a time. The test AUC changes little, both with MKP-Adapter alone and with
\thn, and with MKP-Adapter alone every setting stays far above the plain rank merge.
\tmc is thus robust to its hyperparameters with MKP-Adapter, the backward system tested here.
\section{Forward-compatible backfilling (FCT)}

\begin{figure*}[t]
  \centering
  \includegraphics[width=\linewidth]{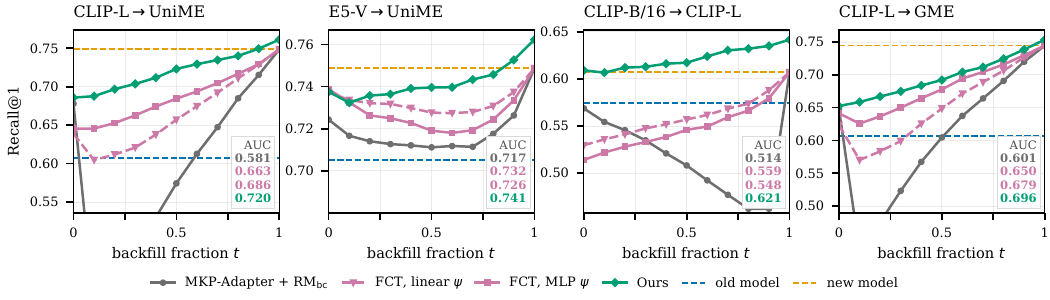}
  \caption{\textbf{Forward-compatible backfilling (FCT) against ours.} Recall@1 over the backfill, averaged over COCO
  and Flickr30K in both directions. FCT~\cite{ramanujan2022fct} (without side information) stores every unmigrated
  item as $\psi(\mathbf{x}^{\text{old}})$, its old embedding mapped into the new space by a forward map $\psi$
  (old to new, the opposite direction of the backward adapter $h$), and searches one index with the new query; $\psi$ is linear (least squares) or an MLP trained by cosine regression, on the same LLaVA-LCS-558K pairs as the backward adapters. Ours: MKP-Adapter trained with \thn, merged with \tmc. Dashed: old and new model alone.
  Box: AUC.}
  \label{fig:appendix_fct}
\end{figure*}

FCT~\cite{ramanujan2022fct} avoids rank merging by mapping embeddings in the opposite direction from backward adapters (\cref{sec:related}).
While a backward adapter $h$ maps new embeddings to the old space, FCT learns a forward map $\psi$ from the old space to the new one.
Before backfilling begins, FCT transforms every gallery item's old embedding $\mathbf{x}^{\text{old}}$ into $\psi(\mathbf{x}^{\text{old}})$.
Queries are then encoded only by the new model and search a single index containing $\psi(\mathbf{x}^{\text{old}})$ for unmigrated items and $\mathbf{x}^{\text{new}}$ for migrated ones.
As backfilling proceeds, the mapped embeddings are replaced by new embeddings item by item, eliminating the need for rank merging.
FCT can also feed $\psi$ side information, an extra feature of each item extracted from the raw data by a separately trained model; this is very expensive, since it needs another encoder pass over every gallery item and another stored vector per item, so we evaluate FCT without side information. Its $\psi$ maps the old embedding alone, either a linear map fitted by least squares or an MLP trained with cosine regression, both on the same embedding pairs as the backward adapters.

It is worth noting that FCT must transform the entire old gallery before the new model
can serve queries (\cref{sec:intro}), a preprocessing cost that grows with the gallery
size. Even with this costly preprocessing, FCT falls well short of \ours.
\Cref{fig:appendix_fct} compares the two on four upgrades on COCO and Flickr30K, the
datasets on which FCT was originally evaluated. \ours achieves a higher AUC on every
upgrade and, on average, a higher Recall@1 at $t=0$.
\section{Random seeds}

\begin{table}[t]
\centering
\small
\setlength{\tabcolsep}{2pt}
\resizebox{\linewidth}{!}{%
\begin{tabular}{l ccccc}
\toprule
Backward system & \shortstack{CLIP-L\\$\rightarrow$UniME} & \shortstack{E5-V\\$\rightarrow$UniME} & \shortstack{CLIP-B/16\\$\rightarrow$CLIP-L} & \shortstack{CLIP-L\\$\rightarrow$GME} & Mean \\
\midrule
RQT~\cite{seo2025mcl} (\rmnaiveh) & 0.528\,{\scriptsize$\pm$0.001} & 0.746\,{\scriptsize$\pm$0.001} & 0.456\,{\scriptsize$\pm$0.002} & 0.530\,{\scriptsize$\pm$0.001} & 0.565\,{\scriptsize$\pm$0.001} \\
\quad + \tmc & 0.702\,{\scriptsize$\pm$0.002} & 0.773\,{\scriptsize$\pm$0.002} & 0.541\,{\scriptsize$\pm$0.001} & 0.680\,{\scriptsize$\pm$0.001} & 0.674\,{\scriptsize$\pm$0.001} \\
\rowcolor{oursgray} \quad + Ours & 0.745\,{\scriptsize$\pm$0.002} & 0.792\,{\scriptsize$\pm$0.002} & 0.629\,{\scriptsize$\pm$0.001} & 0.712\,{\scriptsize$\pm$0.001} & 0.720\,{\scriptsize$\pm$0.001} \\
\midrule
Drift-Adapter~\cite{vejendla2025drift} (\rmnaiveh) & 0.545\,{\scriptsize$\pm$0.001} & 0.734\,{\scriptsize$\pm$0.002} & 0.409\,{\scriptsize$\pm$0.003} & 0.580\,{\scriptsize$\pm$0.002} & 0.567\,{\scriptsize$\pm$0.001} \\
\quad + \tmc & 0.707\,{\scriptsize$\pm$0.001} & 0.772\,{\scriptsize$\pm$0.002} & 0.563\,{\scriptsize$\pm$0.001} & 0.644\,{\scriptsize$\pm$0.001} & 0.672\,{\scriptsize$\pm$0.001} \\
\rowcolor{oursgray} \quad + Ours & 0.746\,{\scriptsize$\pm$0.001} & 0.792\,{\scriptsize$\pm$0.002} & 0.628\,{\scriptsize$\pm$0.001} & 0.704\,{\scriptsize$\pm$0.001} & 0.717\,{\scriptsize$\pm$0.001} \\
\midrule
MKP-Adapter~\cite{byun2026mkp} (\rmnaiveh) & 0.580\,{\scriptsize$\pm$0.001} & 0.783\,{\scriptsize$\pm$0.001} & 0.500\,{\scriptsize$\pm$0.002} & 0.611\,{\scriptsize$\pm$0.002} & 0.619\,{\scriptsize$\pm$0.001} \\
\quad + \tmc & 0.745\,{\scriptsize$\pm$0.001} & 0.790\,{\scriptsize$\pm$0.001} & 0.587\,{\scriptsize$\pm$0.001} & 0.715\,{\scriptsize$\pm$0.001} & 0.709\,{\scriptsize$\pm$0.000} \\
\rowcolor{oursgray} \quad + Ours & 0.757\,{\scriptsize$\pm$0.001} & 0.799\,{\scriptsize$\pm$0.002} & 0.630\,{\scriptsize$\pm$0.001} & 0.723\,{\scriptsize$\pm$0.001} & 0.727\,{\scriptsize$\pm$0.001} \\
\bottomrule
\end{tabular}}
\caption{\textbf{Random seeds.} Recall@1 AUC in the image--text setting of \cref{tab:main}, mean $\pm$ standard
deviation over five random seeds, each drawing both the migration order of the gallery and the arrival order of the
queries (seed 0 is the one of \cref{tab:main}). + \tmc: the backward system as trained, merged with \tmc.}
\label{tab:seeds}
\end{table}

Unless stated otherwise, every result in the paper uses one random migration order of the gallery and one random arrival order of the queries. At each evaluated backfill fraction $t$, the gallery partitions are fixed and the queries are served
one at a time in this order; \tmc's running statistics start anew at each $t$, each query using those of the queries served before it (the first query, its own tails). Repeating the evaluation over five random seeds, each drawing both orders (\cref{tab:seeds}), the standard deviation of the AUC over the seeds is small for every backward system, merged plainly, with \tmc or with
\ours, far smaller than the margins between them.

\section{Non-random migration orders}

\begin{table}[t]
\centering
\small
\setlength{\tabcolsep}{3pt}
\resizebox{\linewidth}{!}{%
\begin{tabular}{l cccc c}
\toprule
Migration order & random & least conf. & clusters & popular & Mean \\
\midrule
\multicolumn{6}{l}{\textbf{MKP-Adapter~\cite{byun2026mkp}}} \\
\quad \rmnaiveh & 0.618 & 0.621 & 0.638 & 0.624 & 0.625 \\
\quad Global map & 0.610 & 0.609 & 0.626 & 0.613 & 0.615 \\
\quad Z-score~\cite{montague2001normalization} & 0.676 & 0.670 & 0.611 & 0.643 & 0.650 \\
\quad Ours (\tmc) & 0.710 & 0.709 & 0.700 & 0.694 & 0.703 \\
\rowcolor{oursgray} \quad Ours & \textbf{0.727} & \textbf{0.727} & \textbf{0.716} & \textbf{0.709} & \textbf{0.720} \\
\midrule
\multicolumn{6}{l}{\textbf{RQT~\cite{seo2025mcl}}} \\
\quad \rmnaiveh & 0.564 & 0.543 & 0.558 & 0.563 & 0.557 \\
\quad Global map & 0.638 & 0.646 & 0.651 & 0.639 & 0.643 \\
\quad Z-score~\cite{montague2001normalization} & 0.649 & 0.645 & 0.586 & 0.620 & 0.625 \\
\quad Ours (\tmc) & 0.673 & 0.681 & 0.665 & 0.659 & 0.669 \\
\rowcolor{oursgray} \quad Ours & \textbf{0.719} & \textbf{0.722} & \textbf{0.707} & \textbf{0.700} & \textbf{0.712} \\
\midrule
\multicolumn{6}{l}{\textbf{Drift-Adapter~\cite{vejendla2025drift}}} \\
\quad \rmnaiveh & 0.566 & 0.565 & 0.571 & 0.569 & 0.567 \\
\quad Global map & 0.566 & 0.553 & 0.577 & 0.563 & 0.565 \\
\quad Z-score~\cite{montague2001normalization} & 0.636 & 0.637 & 0.577 & 0.608 & 0.614 \\
\quad Ours (\tmc) & 0.671 & 0.673 & 0.663 & 0.652 & 0.664 \\
\rowcolor{oursgray} \quad Ours & \textbf{0.716} & \textbf{0.719} & \textbf{0.706} & \textbf{0.697} & \textbf{0.709} \\
\bottomrule
\end{tabular}}
\caption{\textbf{Non-random migration orders.} Recall@1 AUC, mean of the four upgrades and of COCO, Flickr30K and
Urban1k in both directions, for each backward adapter. Global map
and z-score~\cite{montague2001normalization}: the score normalizations of \cref{sec:baselines}. Mean: over the four
orders. Bold: best per column within each block. Confidence: the old model's max softmax probability
($\tau{=}0.05$) of an item over the test queries' old embeddings; least conf.: least confident items first (MCL's and
FastFill's priority); clusters: 10 k-means clusters of the old gallery, migrated one after another;
popular: most often in the old system's top-10 for 8,000 LLaVA-LCS-558K queries first.}
\label{tab:order}
\end{table}

Backfilling may migrate the gallery in a prioritized rather than a random order: MCL~\cite{seo2025mcl} and
FastFill~\cite{jaeckle2023fastfill} prioritize items for which the old model
is least confident, while practical systems may migrate popular items first
or process the index shard by shard. In these cases, the migrated partition
is no longer a uniformly random subset of the gallery.

\Cref{tab:order} compares four migration orders:
\begin{itemize}
    \item \textbf{Random}: Items are migrated in uniformly random order,
    as in the rest of the paper.
    \item \textbf{Least conf.}: Items with the lowest confidence under the
old model are migrated first, following MCL and FastFill.
Without labels, we estimate confidence using only old embeddings,
treating test queries as classes. For each gallery item, we apply
softmax ($\tau=0.05$) to its cosine similarities with all test queries
and take the maximum probability as its confidence. Confidence is
high when one query stands out clearly and low when the item is
similarly close to many queries. Items are migrated in ascending
order of confidence. The test queries are used solely to determine the shared migration
order; no model parameters or global calibration maps are fitted
using them.
    \item \textbf{Clusters}: Old gallery embeddings are partitioned into
    10 clusters using $k$-means. Clusters are migrated in random order,
    with items within each cluster also randomly ordered, simulating
    shard-wise migration.
    \item \textbf{Popular}: Frequently retrieved items are migrated first.
    As the test sets lack query logs, we use 8,000 queries sampled from
    LLaVA-LCS-558K (images for I$\rightarrow$T and captions for
    T$\rightarrow$I), excluding the test queries. Each query is encoded
    by the old model and retrieves its top-10 gallery items by cosine
    similarity. An item's popularity is the number of top-10 lists
    containing it. Items are migrated in descending popularity, with
    ties broken randomly.
\end{itemize}

Averaged across upgrades, Ours achieves the best performance for every
backward adapter under every migration order, consistently outperforming
the raw merge, global map, and z-score normalization~\cite{montague2001normalization}.
Z-score normalization, the strongest baseline under ``random'', is the
most sensitive to migration order, even falling below the raw merge
with MKP-Adapter under ``clusters''.
\section{Gallery size and approximate search}

\begin{table*}[t]
\centering
\small
\setlength{\tabcolsep}{3pt}
\resizebox{\textwidth}{!}{%
\begin{tabular}{l ccccc ccccc}
\toprule
& \multicolumn{5}{c}{exhaustive search} & \multicolumn{5}{c}{IVF search} \\
\cmidrule(lr){2-6}\cmidrule(lr){7-11}
Distractors & \rmnaiveh & Global map & Z-score~\cite{montague2001normalization} & Ours (\tmc) & Ours & \rmnaiveh & Global map & Z-score~\cite{montague2001normalization} & Ours (\tmc) & Ours \\
\midrule
10K & 0.597 & 0.598 & 0.627 & 0.676 & \textbf{0.695} & 0.583 & 0.584 & 0.650 & 0.659 & \textbf{0.676} \\
100K & 0.549 & 0.551 & 0.580 & 0.632 & \textbf{0.656} & 0.541 & 0.541 & 0.607 & 0.621 & \textbf{0.643} \\
Full & 0.482 & 0.488 & 0.520 & 0.562 & \textbf{0.591} & 0.476 & 0.481 & 0.525 & 0.554 & \textbf{0.583} \\
\bottomrule
\end{tabular}}
\caption{\textbf{Gallery size and approximate search.} Recall@1 AUC, mean of the four upgrades, COCO and Flickr30K in
both directions (one query per image for T$\rightarrow$I), MKP-Adapter as the backward system (trained with \thn in
the Ours columns). The test gallery is extended by distractors from LLaVA-LCS-558K (captions for
I$\rightarrow$T, images for T$\rightarrow$I), which migrate with it in a random order. Rows: the number of distractors added (Full: all of LLaVA-LCS-558K except the items used to fit the global maps). IVF: each system's index has $4\sqrt{G}$
k-means lists and probes 1/32 of them; the merge sees only what the index returns, and \tmc reads its tails from it
(z-score normalization~\cite{montague2001normalization} likewise standardizes by the whole list returned from each partition, its top 1000). Global map, z-score:
as in \cref{sec:baselines}.
Bold: best per search setting.}
\label{tab:scale}
\end{table*}
Since our main evaluations use relatively small test galleries,
we further examine the scalability of our method to larger galleries
and approximate nearest-neighbor search.
\Cref{tab:scale} expands the test gallery of each upgrade with
10K, 100K, or all LLaVA-LCS-558K items (except those used
to fit the global maps) as distractors, under two search strategies.
Since the distractors come from the adapters' training corpus, this tests
robustness to gallery size and approximate search, not generalization
to unseen distractors.
\emph{Exhaustive search}, used throughout the rest of the paper,
scores each query against every gallery item, returning exact
top-ranked results.
\emph{IVF search} (inverted file index), a common approximate
search strategy for large galleries, partitions each system's
gallery embeddings into $4\sqrt{G}$ lists using $k$-means
($G$: gallery size). For each query, it searches only the
1/32 of lists whose centroids are closest to the query.
Items outside the probed lists are excluded from both retrieval
results and the score tails used by \tmc. Relevant items
may thus be missed, and tail statistics come from only a subset
of each partition. Each system's index is built once over the whole
extended gallery, and each partition keeps only its own items from
the returned list.

Performance decreases for all methods as distractors are added,
but Ours consistently achieves the best results at every gallery
size under both search strategies.
Under IVF search, most baselines lose some AUC, particularly
on the smallest gallery, where fewer items are retrieved from
the probed lists. In contrast, z-score normalization~\cite{montague2001normalization}
improves under IVF search, as it standardizes scores using only
the items returned by the index.

\section{Score distributions in more detail}

\begin{figure*}[tp]
  \centering
  \vspace*{-4pt}\textbf{(a)}~Homogeneous model, unimodal\par\smallskip
  \includegraphics[width=0.87\linewidth,trim=0 0 0 12,clip]{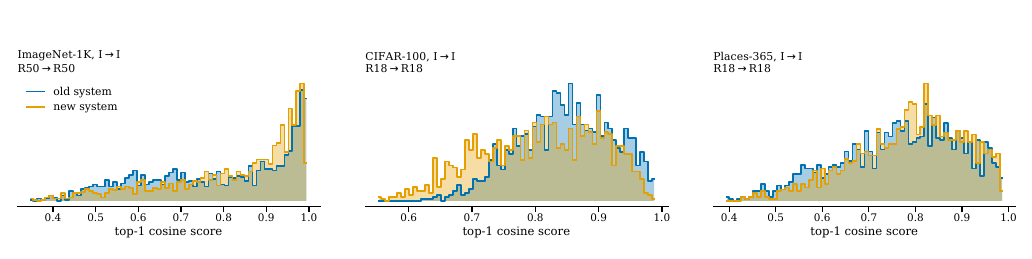}\par
  \includegraphics[width=0.87\linewidth]{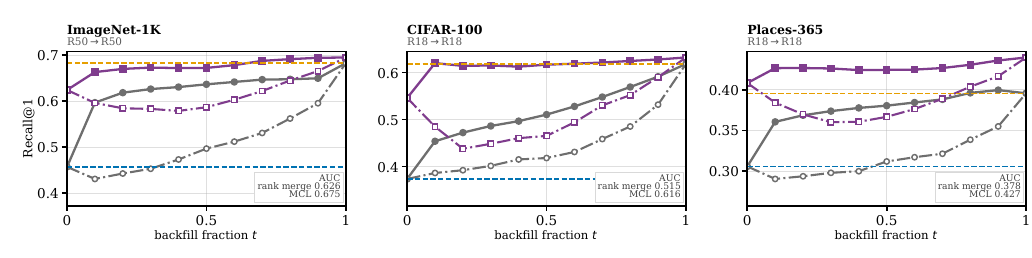}\par\smallskip
  \textbf{(b)}~Heterogeneous model, unimodal\par\smallskip
  \includegraphics[width=0.87\linewidth,trim=0 0 0 12,clip]{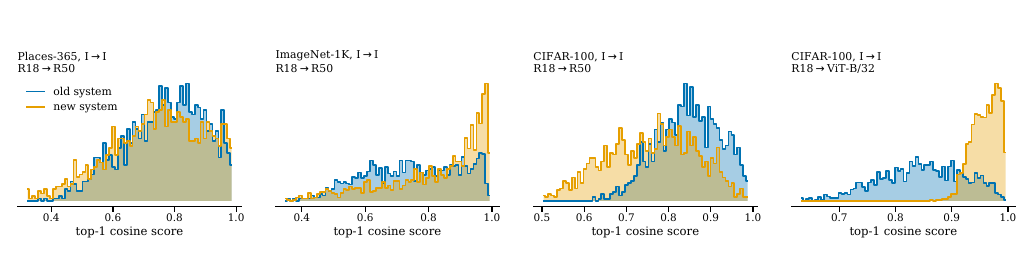}\par
  \includegraphics[width=0.87\linewidth]{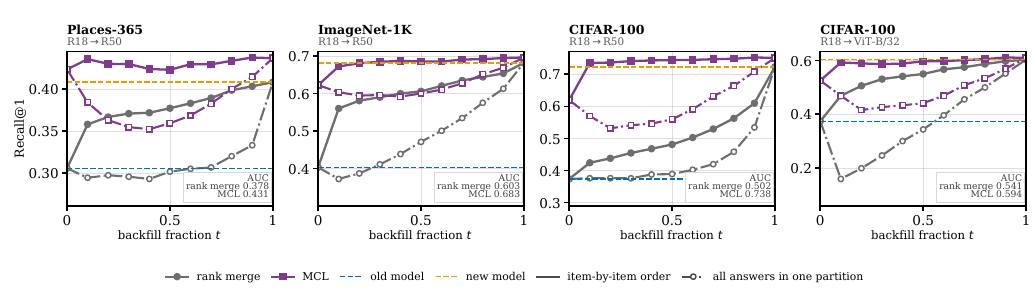}
  \caption{\textbf{MCL's benchmarks, setup by setup} (the seven setups of~\cref{fig:backfill}a; our reproduction,
  \cref{sec:appendix_motivation}). Distributions: each query's top-1 score in the old and in the new system.
  Curves: Recall@1 (top-1
  accuracy with class labels) over the backfill with MCL's old-model-confidence order and a raw merge, for the rank
  merge (\rmnaive) and MCL (our reproduction); dashed: the old and the new model alone; box: area under the curve.
  Solid: MCL's item-by-item order; dash-dot, open markers: the same models with all of a query's correct items in
  one partition, whole classes migrated as blocks in random order.}
  \label{fig:setups_mcl}
\end{figure*}

\begin{figure*}[tp]
  \centering
  \includegraphics[width=\linewidth]{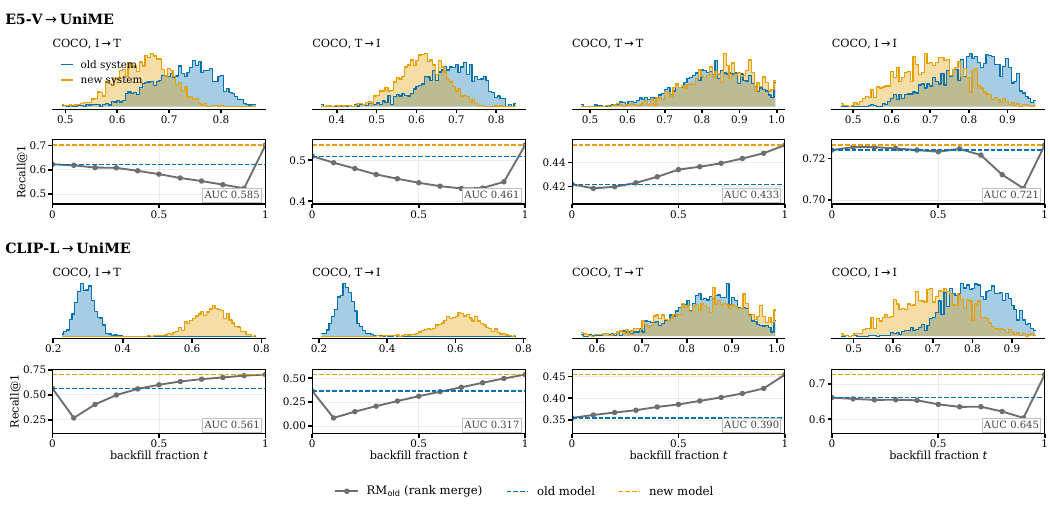}
  \caption{\textbf{What separates the two systems' top-1 distributions, and what it does to the merge.} Image-to-text, text-to-image, text-to-text and image-to-image retrieval on
  COCO for an upgrade within one model family (E5-V$\rightarrow$UniME) and one across families
  (CLIP-L$\rightarrow$UniME). Top: each
  query's top-1 score in the old and the new system. Bottom: Recall@1 of
  \rmnaive over the backfill, random order; box: its AUC. We omit MCL for clarity, as its multimodal extension does not exhibit a consistent trend across these tasks (\cref{sec:appendix_mcl_repro}), which could obscure the comparison. I$\rightarrow$I uses class-level labels: an image is relevant if it shares the query's primary COCO category, the one covering the largest area in the image. Dashed: the old and the new
  model alone.}
  \label{fig:dist_causes}
\end{figure*}

\Cref{fig:setups_mcl} presents setup-wise results from our reproduction
of MCL's benchmarks (\cref{sec:appendix_mcl_repro}).
Under MCL's original protocol (solid lines), the two top-1 score
distributions overlap in every setup except CIFAR-100
R18$\rightarrow$ViT-B/32. The naive rank merge (\rmnaive) outperforms
the old model from the onset of migration, with MCL performing even
better. This protocol is relatively forgiving: under class-level
relevance, each query has 49 (ImageNet-1K) to 99 (CIFAR-100,
Places-365) correct items, and item-wise migration tends
to distribute them across both partitions.

To examine a more challenging setting, we migrate entire classes
as blocks in random order (dash-dot lines). This places all correct
items for a given query in a single partition, as in instance-level
retrieval. Although the endpoints remain unchanged, many more setups
now exhibit a sharp performance drop at the onset of migration,
with both methods falling below the straight line connecting their
endpoints. This behavior closely resembles what we observe in
our realistic retrieval scenarios
(\cref{fig:backfill}b; see also \cref{fig:distributions}c).

\Cref{fig:dist_causes} further examines similarity score distribution
mismatch across modalities and model families on COCO, comparing
an intra-family upgrade (E5-V$\rightarrow$UniME) with an
inter-family upgrade (CLIP-L$\rightarrow$UniME).
First, in cross-modal retrieval (left two columns), the top-1
score distributions exhibit a larger mismatch for the inter-family
upgrade (bottom row) than for the intra-family upgrade (top row),
although each is represented by a single upgrade.
For CLIP-L$\rightarrow$UniME, the distributions are nearly disjoint,
with the new model consistently producing higher top-1 scores,
whereas E5-V$\rightarrow$UniME exhibits a smaller shift.
Second, for the inter-family upgrade, the mismatch is much more
pronounced in cross-modal than in unimodal retrieval (right two
columns). For the intra-family upgrade, however, image-to-image
retrieval exhibits a shift comparable to that of cross-modal
retrieval. Notably, in text-to-text retrieval, the distributions nearly
coincide for both upgrades, allowing \rmnaive to improve
almost linearly from the old to the new model, as in MCL's
original setups (\cref{fig:backfill}a).
\section{MCL reproduction}
\label{sec:appendix_mcl_repro}
\label{sec:appendix_motivation}

\nbf{MCL's benchmarks}
\Cref{fig:backfill}a and \cref{fig:setups_mcl} reproduce seven setups of~\cite{seo2025mcl} on ImageNet-1K, CIFAR-100
and Places-365 with MCL's default settings: as in~\cite{seo2025mcl}, MCL is trained with class labels on the training
set of each benchmark and evaluated in Recall@1 (CMC@1) with its old-model-confidence migration order. Our reproduction closely matches the reported results for both
\rmnaive and MCL across most setups (within 2.7 AUC points), with a larger discrepancy
for MCL on CIFAR-100 R18$\rightarrow$R50, where ours is 9.3 points higher.
Conversely, our reproduction of MCL's Market-1501 setups yielded substantially
lower results than those reported.
These discrepancies may stem from backbone-specific configurations that are
not fully specified in the original paper.
To avoid comparing against potentially unrepresentative baselines, we omit MCL's Market-1501 setups.

\nbf{MCL in our setups}
Since image--caption datasets lack class labels, we treat each
image--caption pair as a separate class, using the paired item
as the positive and all other items in the batch as negatives.
LLaVA-LCS-558K, on which MCL is trained, pairs each image with one
caption, so each anchor has exactly one positive.
We otherwise follow MCL's original method and default settings.
For cross-modal retrieval, we train MCL bidirectionally, using
images as anchors against captions (I$\rightarrow$T) and captions
as anchors against images (T$\rightarrow$I), summing the two losses.

\nbf{Why MCL underperforms in our setups}
Unlike on its original benchmarks, MCL falls well below \ours
on our upgrades (\cref{tab:main}).
Moreover, its performance relative to the naive rank merge
(\rmnaive) shows no consistent trend across tasks, test sets,
or upgrades: MCL outperforms \rmnaive in some cases (mostly
T$\rightarrow$I) but underperforms it in others (mostly
T$\rightarrow$T).
We attribute this to two possible factors.
First, our implementation is a straightforward multimodal extension
of MCL, which was originally designed for unimodal retrieval.
Extending it to cross-modal retrieval may require a more sophisticated
approach than simply summing bidirectional (I$\rightarrow$T and
T$\rightarrow$I) contrastive losses.
More importantly, MCL's original benchmarks use in-domain training
data, whereas our experiments train on a separate image--caption
corpus and evaluate on COCO, Flickr30K, and Urban1k.
This domain gap may be a major factor behind its weaker and less
consistent performance.
\end{document}